\documentclass[aps,rmp,reprint,amsmath,amssymb,graphicx,longbibliography]{revtex4-1}

\usepackage[utf8]{inputenc}
\usepackage{bm}
\usepackage{graphicx}
\usepackage{epstopdf}
\usepackage{wrapfig}
\usepackage{array}
\usepackage{listings}
\usepackage[para,online,flushleft]{threeparttablex}
\usepackage{booktabs,dcolumn}
\usepackage{color}
\usepackage{textpos}
\usepackage{booktabs}
\usepackage{multirow,bigdelim}
\usepackage{float}
\usepackage{upgreek} 
\usepackage{hyperref}
\usepackage{xcolor}
\usepackage{comment}
\usepackage{physics}
\usepackage{braket}
\usepackage{environ}
\usepackage{bbm}
\usepackage{xspace}
\usepackage{isotope}

\newcolumntype{C}[1]{>{\centering\arraybackslash}p{#1}}

\hypersetup{breaklinks=true,colorlinks=true,linkcolor=blue,citecolor=blue,
filecolor=magenta,urlcolor=blue}

\makeatletter
\def\@bibdataout@aps{%
\immediate\write\@bibdataout{%
@CONTROL{%
apsrev41Control%
\longbibliography@sw{%
    ,author="08",editor="1",pages="1",title="0",year="1"%
    }{%
    ,author="08",editor="1",pages="1",title="",year="1"%
    }%
  }%
}%
\if@filesw \immediate \write \@auxout {\string \citation {apsrev41Control}}\fi
}
\makeatother

\newcommand{\param}{\theta}
\newcommand{\params}{\boldsymbol{\theta}}
\newcommand{\target}{*}
\newcommand{\nb}{n_b}
\newcommand{\wt}{\widetilde}
\newcommand{\betavec}{\vec\beta}

\NewEnviron{subalign}[1][]{%
\begin{subequations}\begin{align}
  \BODY
\end{align}\label{#1}\end{subequations}
}

\NewEnviron{spliteq}{%
\begin{equation}\begin{split}
  \BODY
\end{split}\end{equation}
}

\newcommand{\abinitio}{\textit{ab initio}\xspace}
\newcommand{\chieft}{\ensuremath{\chi}EFT\xspace}
\newcommand{\NEC}{\ensuremath{n_b}}
\newcommand{\Nmax}{\ensuremath{N_\text{max}}}

\begin{document}

\title{Eigenvector Continuation and Projection-Based
Emulators}

\author{Thomas~Duguet}
\affiliation{IRFU, CEA, Universit{\'e} Paris-Saclay, 91191 Gif-sur-Yvette, France}
\affiliation{KU Leuven, Department of Physics and Astronomy, Instituut voor Kern- en Stralingsfysica, 3001 Leuven, Belgium}

\author{Andreas~Ekstr{\"o}m}
\affiliation{Department of Physics, Chalmers University of Technology, SE-412 96 G{\"o}teborg, Sweden}

\author{Richard~J.~Furnstahl}
\affiliation{Department of Physics, The Ohio State University, Columbus, OH 43210, USA}

\author{Sebastian~K{\"o}nig}
\affiliation{Department of Physics, North Carolina State University, Raleigh, NC 27695, USA}

\author{Dean~Lee}
\affiliation{Facility for Rare Isotope Beams \& Department of Physics and Astronomy, Michigan State University, MI 48824, USA}

\begin{abstract}
Eigenvector continuation is a computational method for parametric eigenvalue problems that uses subspace projection with a basis derived from eigenvector snapshots from different parameter sets.  It is part of a broader class of subspace-projection techniques called reduced-basis methods.  In this colloquium article, we present the development, theory, and applications of eigenvector continuation and projection-based emulators.  We introduce the basic concepts, discuss the underlying theory and convergence properties, and present recent applications for quantum systems and future prospects.

\end{abstract}

\maketitle

\tableofcontents

\section{Motivation}\label{sec:motivation}

The challenges of nuclear few- and many-body physics have been addressed theoretically with a wide range of accurate 
but often
computationally expensive \emph{high-fidelity} methods.
However, when we need to change the parameters characterizing the problem, such as Hamiltonian coupling constants, it can become computationally prohibitive to repeat high-fidelity calculations many times and challenging to reliably extrapolate.
An alternative is to replace the high-fidelity model with an \emph{emulator}, which is an approximate computer model, in the literature sometimes referred to as a ``surrogate.''
We focus in this colloquium on the recent development and application of emulators that exploit a technique called eigenvector continuation (EC) and its extensions.
Our illustrative examples are primarily drawn from nuclear structure and reactions, for which there has been an explosion of EC applications in the last few years.  We emphasize, however, the general scope of the methods, which are broadly applicable to physics problems.

\begin{figure}[tbp] \centering
 \includegraphics[width=1.0\columnwidth]{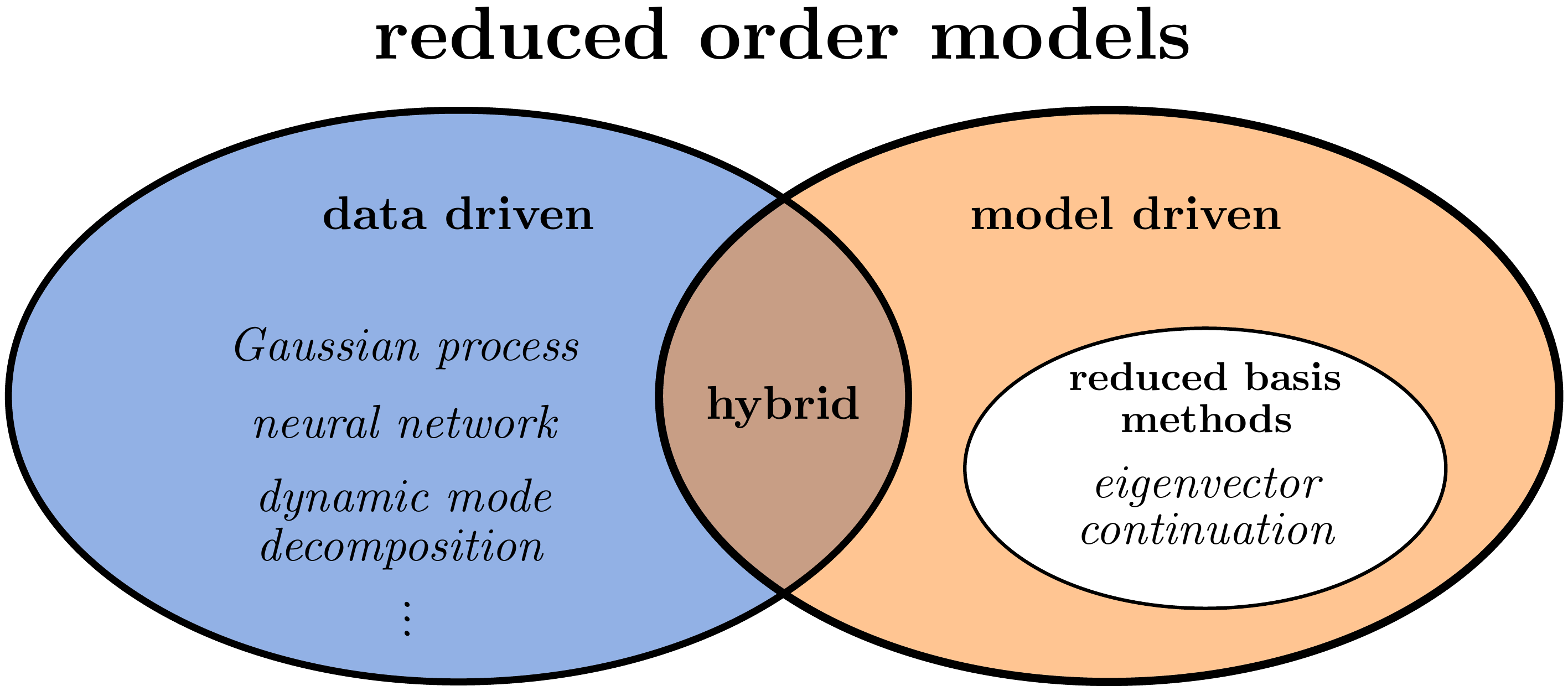}
 \caption{%
  Schematic classification of model order reduction emulators into data-driven methods, including Gaussian processes,  artificial neural networks, and dynamic mode decomposition; model-driven methods, including reduced-basis methods (RBMs); and hybrid methods. Eigenvector continuation (EC) approaches are a subset of RBM.
  \label{fig:MOR_venn_diagram}
 }
\end{figure}

Even more broadly, being able to efficiently vary the parameters in high-fidelity models to enable design, control, optimization, inference, and uncertainty quantification is a general need across engineering and science fields~\cite{Benner2020Volume1DataDriven,Benner20201,Benner2020Volume3Applications}.
A common theme in these endeavors is that much of the information in high-fidelity models is superfluous. This can be exploited when tracing parametric dependencies by reducing the complexity through a so-called reduced order model, i.e., an emulator.
The universe of model order reduction methods is relatively mature but continues to expand, along with their applications.

We can put EC emulators in perspective by considering 
a high-level classification of reduced order models into data-driven and model-driven categories (see Fig.~\ref{fig:MOR_venn_diagram}).
Data-driven methods typically interpolate the outputs of high-fidelity models without requiring an understanding of the underlying model structure; examples include Gaussian processes, artificial neural networks, and dynamic mode decomposition.
Model-driven methods solve reduced-order equations derived from the full equations, so they are physics-based and respect the underlying structure; examples include the broad class of reduced-basis methods or RBMs (\textcite{hesthaven2015certified,Quarteroni:218966}).
Increasingly, there are hybrid approaches drawing from knowledge about the underlying physics problem and thereby combining both data- and model-driven aspects (e.g.~\textcite{chen2021physics}).

Although originally developed independently, 
the model-driven EC method has long-established antecedents among RBMs (e.g., eigenvalue problems in structural engineering in \textcite{Aktas:1998,Nair1998} and applied mathematics in \textcite{Machiels2000}, with more recent applications in \textcite{Fumagalli2016,Horger2017,pichi2020reduced}).
EC uses a basis derived from selected eigenvectors from different parameter sets, called \emph{snapshots} in the RBM world, to project into a much smaller subspace than the original problem.
In its simplest form, EC generates a highly effective variational basis.
Typically, EC applications exploit the RBM \emph{offline}-\emph{online} workflow, in which expensive high-fidelity calculations are performed once in the offline phase, enabling inexpensive but still accurate emulator calculations in the online phase.

When the offline-online workflow is applied to calculate observables for many parameters characterizing Hamiltonians or other operators, the EC emulators can achieve tremendous speed-ups over high-fidelity computational methods. This facilitates large-scale parameter exploration and calibration as well as uncertainty quantification, sensitivity analysis, and experimental design that would otherwise be infeasible.
The model-driven nature of the EC approach ensures not only accurate interpolation in the parameter space, but in many cases provides accurate \emph{extrapolations} in the spaces of control parameters such as coupling strengths, energies, and boundary conditions.
A consequence is that problems that are difficult or even intractable for some range of control parameters can be attacked by calculating in a range that can be more easily solved, and then extrapolating using the emulator.

Reliable emulator technology is also useful for collaboration as it enables the development of self-contained and accurate mini-applications that mimic the output of complex model calculations~\cite{Zhang:2021jmi,Tews:2022yfb}. These emulators are easy to distribute given their typically small memory footprint. This allows other researchers to generate fast and accurate model predictions, even without the in-depth knowledge and significant computational resources typically required to create applications from complex or closed-source codebases.

In Sec.~\ref{sec:background} we review the basic concepts of EC and the early work in nuclear physics. A brief overview of the RBM formulation and the offline-online workflow is given in Sec.~\ref{sec:MOR}, along with alternative approaches to generalized eigenvalue problems from a nuclear physics perspective.
EC convergence properties are covered in
Sec.~\ref{sec:theory}, including the application to many-body perturbation theory. 
A major EC application is to large Hamiltonian eigensystems (Sec.~\ref{sec:sparse}), which include adaptations to the shell model and the coupled cluster method.
As illustrations of the wide range of EC applications, extensions are described for scattering (Sec.~\ref{sec:scattering}), finite volume dependence and resonances (Sec.~\ref{sec:volume}), and quantum Monte Carlo simulations (Sec.~\ref{sec:QMC}). A summary and consideration of future directions are presented in Sec.~\ref{sec:future}.

\section{Background}\label{sec:background}

The development of EC in \textcite{Frame:2017fah} was inspired by the quantum many-body problem and the desire to
find the extremal eigenvalues and eigenvectors of a Hamiltonian matrix too large to store in computer memory.  While there are numerous quantum many-body methods used to solve such problems, they all fail when some control parameter in the Hamiltonian matrix exceeds some threshold value.  Monte Carlo methods break down when there are strong sign oscillations and positive and negative amplitudes cancel.  Simple order-by-order summations of diagrammatic expansions and perturbation theory are divergent when the magnitude of the expansion parameter exceeds unity.  Variational methods are not effective if there are strong correlations not adequately captured by some wave-function ansatz or truncated set of basis states.

In the following, we review some of the concepts of EC as well as the early literature.  Let us consider a family of matrix Hamiltonians $H(\params)$ that depends analytically on some vector of control parameters $\params$, which we write in vector notation.  We assume that the matrix Hamiltonians are Hermitian for all real values of the parameters.  One particularly interesting and important example is the affine case where the dependence on each parameter decomposes as a sum of terms
\begin{equation}
    H(\params) = \sum_\alpha f_\alpha(\params) H_\alpha,
    \label{eq:affine}
\end{equation}
for some functions $f_\alpha$ and Hermitian matrices $H_\alpha$. 
We will be interested in the properties of some particular eigenvector of $H(\params)$ and its corresponding eigenvalue $E(\params)$, 
\begin{equation}
    H(\params)\ket{\psi(\params)} = E(\params)\ket{\psi(\params)}.
\end{equation}
The basic idea of eigenvector continuation is that $\ket{\psi(\params)}$ is an analytic function for real values $\params$, and the smoothness implies that it approximately lies on a linear subspace with a finite number of dimensions.  We note that if there are exact eigenvalue degeneracies, the relative ordering of eigenvalues may change as we vary $\params$.  However, the eigenvectors can still be defined as analytic functions in the neighborhood of these exact level crossings. The smoother and more gradual the undulations in the eigenvectors, the fewer dimensions needed.  A good approximation to $\ket{\psi(\params)}$ can be found efficiently using a variational subspace composed of snapshots of $\ket{\psi(\params_i)}$ for parameter values $\params_i$.  We note that for complex values of the parameters, the guarantee of smoothness no longer holds.

At this point we note that other methods exist that are
based on projecting a large-scale linear algebra problem into a low-dimensional subspace.
Krylov methods, and in particular Lanczos/Arnoldi iteration for calculating extremal eigenvalues of linear operators, are well established (see for example \textcite{Saad:2011} for an excellent textbook discussion) and very broadly used, not only in Physics.
An important distinction compared to EC is, however, that these Krylov methods are employed at fixed $\params$, and thus they solve a much
more limited problem.
In fact, many of the EC applications discussed in Sec.~\ref{sec:sparse} would typically use Lanczos iteration to determine the individual $\ket{\psi(\params)}$ snapshots for the EC offline stage.

\begin{figure}[b]
 \includegraphics[width=0.98
\columnwidth]{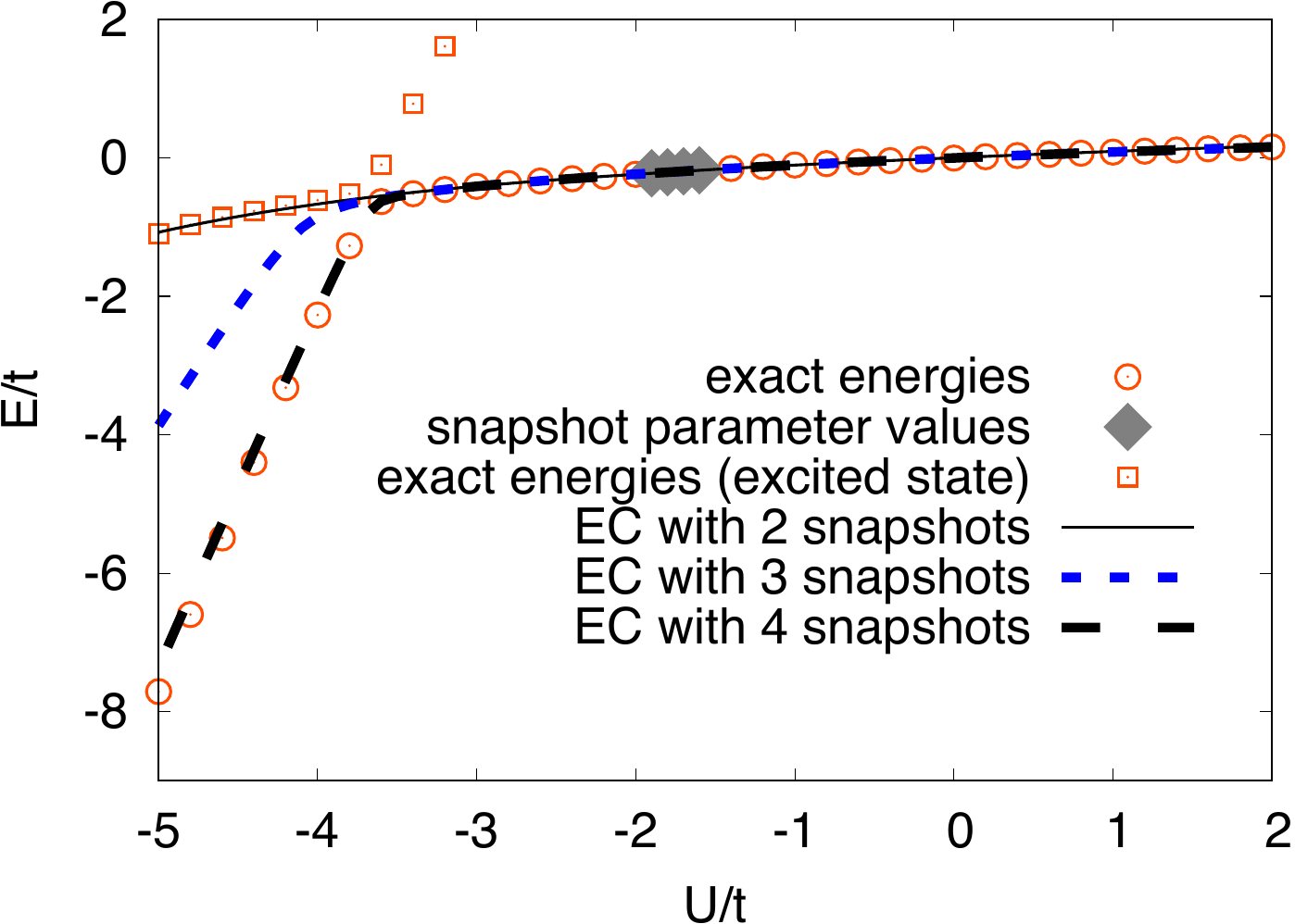}
 \caption{Ground state energy $E$ of the Bose-Hubbard model divided by $t$ versus $U/t$. The exact ground-state energies are shown with open circles, while the EC results are shown for variational subspace dimensions varying from $2$ to $4$.  In order to highlight the avoided level crossing, the exact excited state energies are also shown as open squares.   Adapted from \textcite{Frame:2017fah}.
  \label{fig:fourbody_EC}
 }
\end{figure}

Following \textcite{Frame:2017fah}, we consider the Bose-Hubbard model for identical bosons on a three-dimensional cubic lattice 
as an illuminating and pedagogical example of EC, specializing to four bosons on a $4\times 4 \times 4$ spatial lattice.  
The parameter $t$ is the coefficient for the kinetic energy, and the parameter $U$ is the coefficient for the pointlike interaction between pairs of bosons.  For this system the relevant control parameter is the dimensionless ratio $\param = U/t$.  A variational subspace is constructed from snapshots of the eigenvectors for selected training parameters $\param_j$.  With the shorthand $\ket{\psi_j} = \ket{\psi(\param_j)}$, the norm matrix $\wt N_{ij}$ and projected Hamiltonian matrix $\wt H_{ij}(\param)$ are given by
\begin{gather}
    \wt N_{ij} = \braket{\psi_i | \psi_j }, \\
    \wt H_{ij}(\param) = \braket{\psi_i |H(\param)| \psi_j }.
\end{gather}    
The generalized eigenvalue problem is then solved as discussed in Sec.~\ref{sec:MOR}.  While ``norm matrix'' is the name commonly used in the nuclear physics literature for the matrix of inner products between vectors,  we should note that this matrix is called ``Gram matrix'' in the standard mathematical literature.  

In Fig.~\ref{fig:fourbody_EC} we show the ground-state energy $E$ divided by $t$ versus $U/t$, along with an excited state.  
The exact ground state energies are shown with open circles, which reveal a sharp bend near $U/t = -3.8$. 
The sharp bend is caused by an avoided level crossing of eigenvalues, and the abruptness of the bend indicates that there are branch points located near the real axis.
EC results are shown for subspace dimensions varying from $2$ to $4$.  With snapshot parameter values at $U/t = -2.0, -1.9, -1.8, -1.7$, the EC calculation is capable of extrapolating past the sharp bend.  

We can understand how eigenvector continuation is able to extrapolate in this case by
exploring the connection with analytic continuation.  Let us consider a power series expansion of the eigenvector $\ket{\psi(\param)}$ around $\theta = 0$, 
\begin{equation}
 \ket{\psi(\param)} = \lim_{M \rightarrow \infty} \sum_{m=0}^M
 \ket{\psi^{(m)}(0)} \frac{\param^m}{m!}.
 \label{eqn:series}
\end{equation}
When the series converges, we can approximate $\ket{\psi(\param)}$ to any desired accuracy as a finite sum of $M+1$ vectors $\ket{\psi^{(m)}(0)}$ with $m$ ranging from $0$ to $M$.  

\begin{figure}
 \includegraphics[width=0.95
\columnwidth]{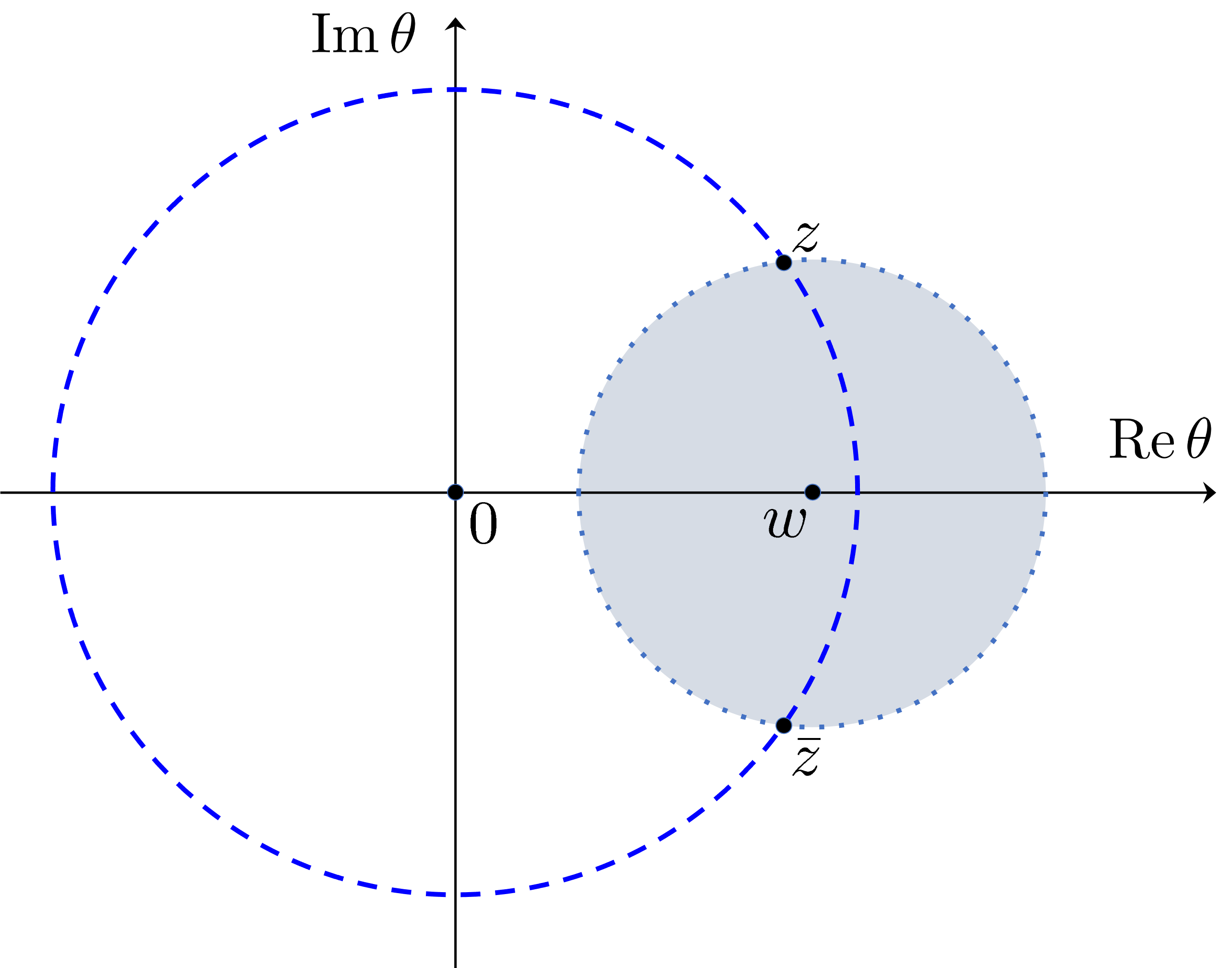}
 \caption{While the power series expansion at $\param=0$ converges only for $|\theta|<|z|$, we can choose a secondary point $w$ with $|w|<|z|$.  The power series expansion at $\param = w$ converges in the shaded region shown and can be re-expressed as a double series around $\theta=0$. \cite{Frame:2017fah}.
  \label{fig:complex_plane}
 }
\end{figure}

The series will diverge when $|\theta|$ exceeds the magnitude of the nearest non-analytic point.  If $H(\theta)$ is a finite-dimensional matrix that depends analytically on $\theta$, then the non-analytic behavior is associated with branch points where two or more eigenvectors become the same vector \cite{kato2013perturbation}.  If $H(\theta)$ is a Hermitian matrix for real $\theta$, then all of the branch points lie away from the real axis and come in complex conjugate pairs.   In Fig.~\ref{fig:complex_plane} we show an example where $z$ and $\bar{z}$ are the nearest branch points to the origin.  While the power series expansion around $\theta=0$ converges only for $|\theta|<|z|$, we can choose a secondary point $w$ with $|w|<|z|$ and expand around $w$,  
\begin{equation}
 \ket{\psi(\param)} = \lim_{N \rightarrow \infty} \sum_{n=0}^N
 \ket{\psi^{(n)}(w)} \frac{(\param-w)^n}{n!}.
\end{equation}
The derivatives at $w$ can in turn be expanded using power series about the origin.  This yields the double sum,  
 \begin{equation}
  \ket{\psi(\param)} =
  \lim_{N \rightarrow \infty} \sum_{n=0}^N \lim_{M \rightarrow \infty} \sum_{m=0}^M
 \ket{\psi^{(n+m)}(0)} \frac{w^m(\param-w)^n}{m! \; n!}.
 \label{eqn:double_sum}
 \end{equation}
We can now approximate $\ket{\psi(\param)}$ in the shaded region in Fig.~\ref{fig:complex_plane} to any desired accuracy as a finite sum of $N+M+1$ vectors $\ket{\psi^{(n+m)}(0)}$, with $n$ ranging from $0$ to $N$ and $m$ ranging from $0$ to $M$.  This process of analytic continuation shows that the approximation of $\ket{\psi(\param)}$ using a finite linear subspace can extend beyond the nearest branch point.  By including additional expansion points, this can be extended to all values of $\param$ where $\ket{\psi(\param)}$ is analytic.

We now return to the general problem where $H(\params)$ depends on a vector of control parameters $\params$.  When the eigenvector snapshots $\ket{\psi(\params_i)}$ are chosen with $\params_i$ infinitesimally close to some common limit point $\bar{\params}$, the variational subspace is spanned by gradients and higher-order gradients of $\ket{\psi(\params)}$ at $\bar{\params}$. The EC calculation is then equivalent to a variational subspace calculation with basis states 
\begin{equation}
\nabla_{i_1} \nabla_{i_2} \cdots \left. \nabla_{i_n}\ket{\psi(\params)} \right|_{\params = \bar{\params}}. \label{eqn:gradients}
\end{equation}
These are the same terms that appear in the perturbation theory expansion of the eigenstate wave function~\cite{Frame:2017fah}. The difference is that we are performing a variational calculation rather than evaluating partial sums of a power series.
In \textcite{Demol:2019yjt}, eigenvector continuation is used to accelerate the convergence of Bogoliubov many-body perturbation theory; this is discussed in Section~\ref{sec:MBPT}. 

The application of EC to quantum Monte Carlo simulations is considered in detail in \textcite{Frame:2019jsw}.  Since quantum Monte Carlo simulations use the Euclidean time evolution operator, $e^{-H(\params)t}$, one produces eigenstates $\ket{\psi(\params)}$ together with exponential factors of $e^{-E(\params)t}$. This produces some technical challenges in applications to large quantum many-body systems.  The resolution of such problems is described in Section~\ref{sec:QMC}.

In \textcite{Konig:2019adq}, it was realized that EC could be used as a fast and accurate emulator for quantum many-body calculations by taking relatively few snapshots $\params_i$ to cover a compact domain of the parameter space.  In \textcite{Ekstrom:2019lss}, the use of EC as an emulator was extended to non-Hermitian matrices as encountered in coupled cluster calculations.  This is discussed in Section~\ref{sec:cc}.  Applications of EC emulators for quantum scattering problems in nuclear physics were first explored in \textcite{Furnstahl:2020abp}.  This work and several subsequent works extending the method and improving the performance are discussed in Section~\ref{sec:scattering}.

As noted in \textcite{Bonilla:2022rph} and \textcite{Melendez:2022kid}, EC should be considered as a special case of a more general area of RBMs, which has been well developed in applied mathematics over several decades, especially in the area of partial differential equations, see for example, \textcite{hesthaven2015certified,Quarteroni:218966}.  Although the early development of EC emphasized quantum many-body systems and extrapolations for eigenvalue problems where the eigenvectors are too large to store in memory, the use of EC as an emulator is very much in line with many other applications of RBMs.

\section{Reduced Basis Methods}\label{sec:MOR}

The literature on RBMs is extensive~\cite{Benner2020Volume1DataDriven,Benner20201,Benner2020Volume3Applications}, with 
recent guides from the perspective of nuclear physicists (and EC), including pedagogical code examples, given in~\textcite{Melendez:2022kid,Drischler:2022ipa}.
Here we touch upon some key features common to EC applications.

\subsection{RBM workflow for a Hamiltonian eigenvalue problem}\label{subsec:workflow}
\begin{figure}[tbp] \centering
 \includegraphics[width=0.9\columnwidth]{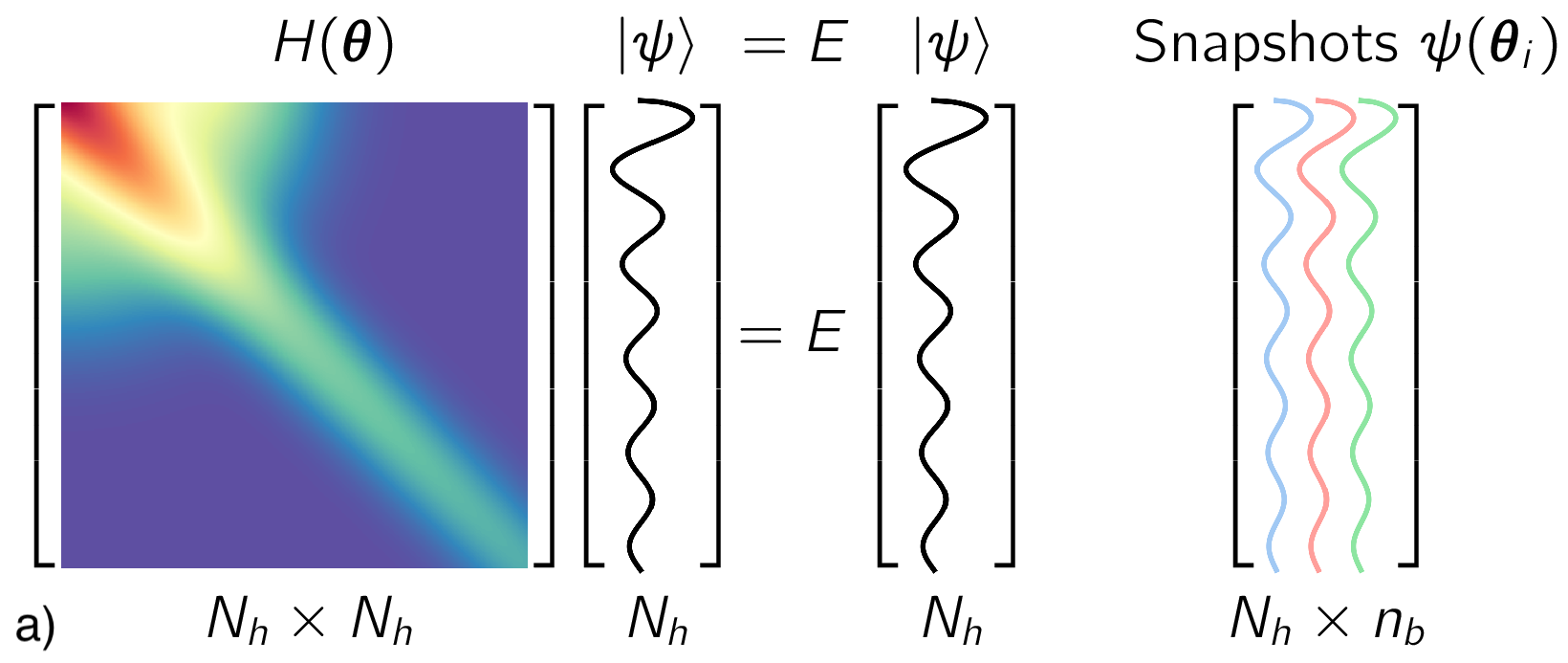}

\rule[0.6ex]{\columnwidth}{0.3mm}

 \includegraphics[width=0.9\columnwidth]{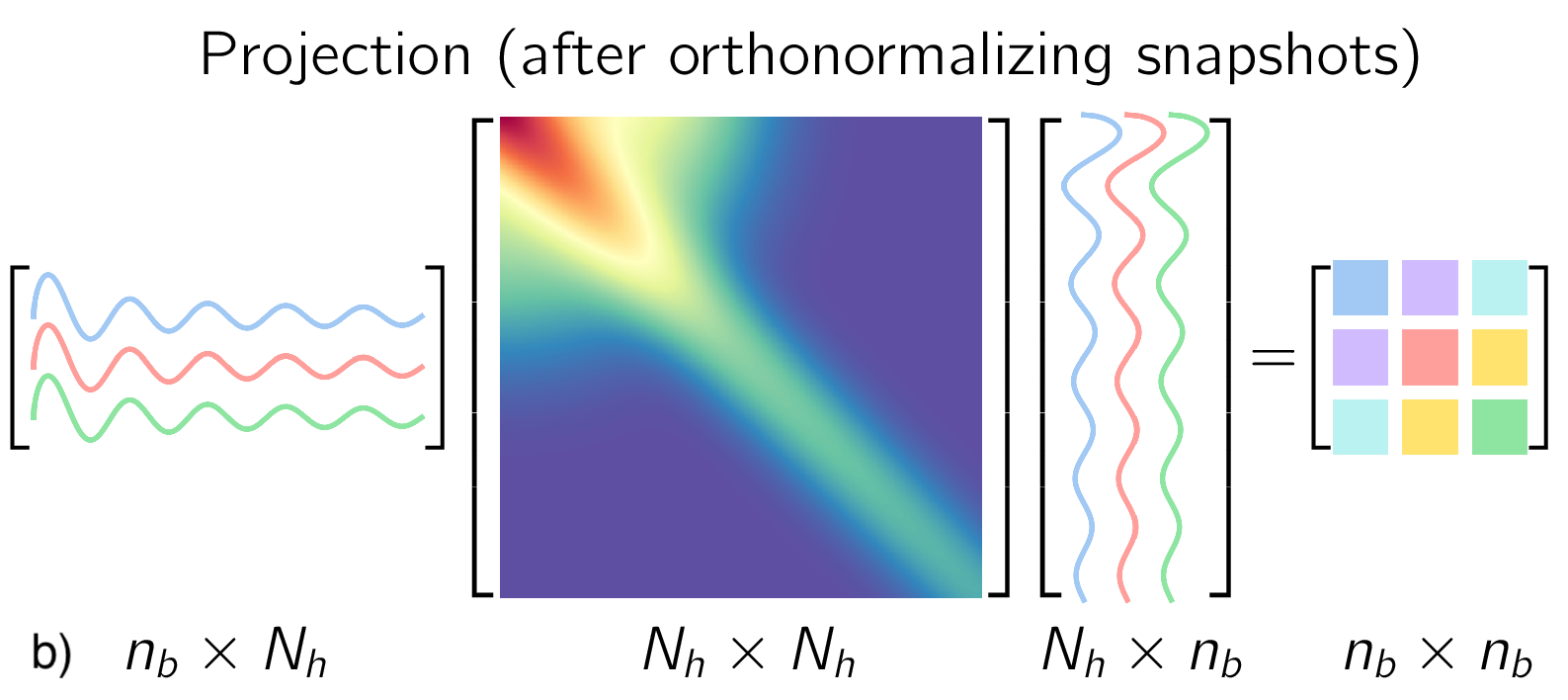}

\rule[0.6ex]{\columnwidth}{0.3mm}

 \includegraphics[width=0.5
\columnwidth]{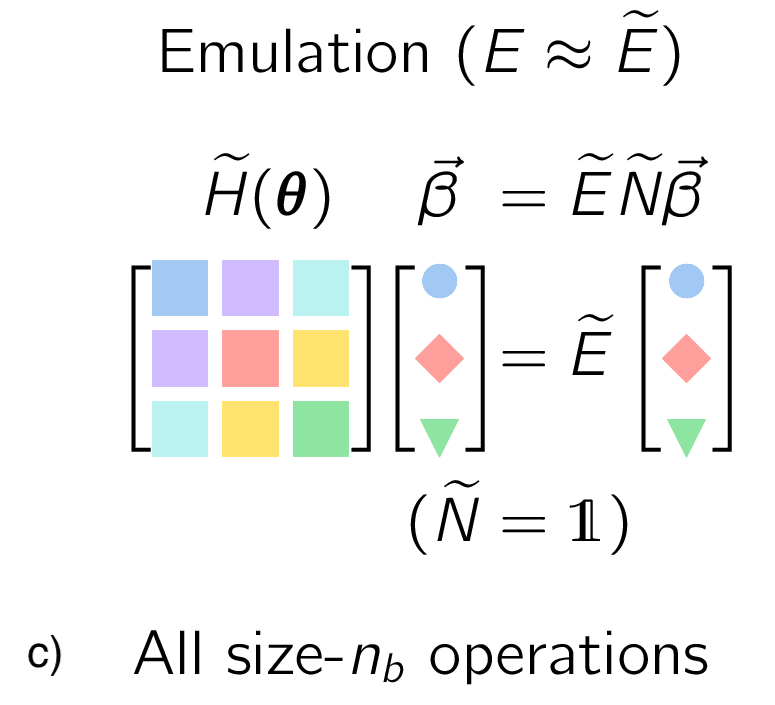}
 \caption{%
 Reduced-basis model workflow for a matrix eigenvalue problem. a) High-fidelity calculations of snapshots, each of large size $N_h$, are b) projected in the \emph{offline stage} to a reduced-basis matrix of small size $n_b\times n_b$. In the c) \emph{online stage}, the emulator only uses size $n_b$ operations. Adapted from a figure in \textcite{Drischler:2022ipa}.
  \label{fig:RBM_offline_online}
 }
\end{figure}

The basic ingredients of an RBM workflow, which is built on a separation into offline and online stages, are illustrated for a familiar Hamiltonian eigenvalue problem in Fig.~\ref{fig:RBM_offline_online}.

\textbf{Formulation in integral form.}
First we cast the equations for the Schr\"odinger wave function or other quantities of interest (such as a scattering matrix) in integral form. 
For the Hamiltonian eigenvalue problem in Fig.~\ref{fig:RBM_offline_online}a (left) with parameters $\params$, solving the finite matrix problem of a (large) basis size $N_h\times N_h$ is formally equivalent to finding $N_h$ (approximate) stationary solutions to the variational functional 
\begin{equation}
      \mathcal{E}[\psi] = \mel{\psi}{H(\params)}{\psi} - E(\params)\bigl(\braket{\psi|\psi} - 1\bigr) ,\label{eq:eigen_variational}
\end{equation}
in the space spanned by the $N_h$ basis elements.
This is our high-fidelity model.
(In practice, the use of Krylov methods mentioned earlier means that finding all $N_h$ solutions is not needed, but the computational cost still scales with $N_h$.)
Other RBM formulations are discussed in Sec.~\ref{subsec:other_formulations}.

\textbf{Offline stage.}
Next we reduce the dimensionality of the problem by substituting for the general solution a trial basis of size $\nb$. RBMs  start with a \emph{snapshot} basis, consisting of high-fidelity solutions $\ket{\psi_i}$ at selected values $\{\params_i;\, i=1,\ldots,n_b\}$ in the parameter space, as in Fig.~\ref{fig:RBM_offline_online}a (right).
When seeking the ground state energy and wave function for arbitrary $\params$, these $\ket{\psi_i}$ are ground-state eigenvectors from diagonalizing $H(\params_i)$.
For many EC applications in nuclear physics to date it has been sufficient to choose this basis randomly, e.g., with a space-filling sampling algorithm such as Latin hypercube sampling. 
This basis spans a reduced space and can be used directly (after orthonormalizing the snapshots),
\begin{equation}
    \ket{\wt\psi} = \sum_{i=1}^{\nb} \beta_i \ket{\psi_i} ,
    \label{eq:trial_basis}
\end{equation}
with basis expansion coefficients $\beta_i$.
The Hamiltonian is then projected to a much smaller $\nb\times\nb$ space, as shown in Fig.~\ref{fig:RBM_offline_online}b.

More generally in RBM applications, one first compresses the snapshot basis by applying some variation of principle component analysis (known as proper orthogonal decomposition or POD in this context), which builds on the singular value decomposition (SVD) of the snapshots and enables a smaller basis size than $n_b$.
Alternatively, or in conjunction with POD, one can efficiently select snapshots by applying an active learning protocol (greedy algorithm) that aims to minimize the overall error of the emulator~\cite{hesthaven2015certified}. For a recent application of a greedy algorithm to quantum spin systems, including an efficient mapping of ground-state phase diagrams, see~\textcite{Herbst2022,Brehmer2023}.
These approaches are the standard in RBM applications, but are not yet widely applied in nuclear physics~\cite{Sarkar:2021fpz,Bonilla:2022rph}.

\textbf{Online stage.}
For variational formulations, we enforce stationarity with respect to the trial basis expansion coefficients. This leads to a $\nb\times\nb$ generalized eigenvalue problem for the basis coefficients,
\begin{align}
    \wt H(\params)\betavec(\params) &= \wt E(\params)\wt N(\params) \betavec(\params) , \notag \\
    \wt H_{ij}(\params) &= \mel{\psi_i}{H(\params)}{\psi_j}, \notag \\
    \wt N_{ij}(\params) &= \braket{\psi_i|\psi_j}  ,
    \label{eq:gen_eigen}
\end{align}
as already introduced in Sec.~\ref{sec:background} and visualized in Fig.~\ref{fig:RBM_offline_online}c.
Note that if the basis has been orthonormalized, then $\wt N$ is an identity matrix.
Extending such an emulator to matrix elements of other operators and even transitions is straightforward; see \textcite{Wesolowski:2021cni} for a nuclear example.

In uncertainty quantification, for which sampling of very many parameter sets are usually required, it is essential that the emulator be many times faster than the high-fidelity calculations.
This is achieved for RBM emulators by the offline and online separation because the online stage only requires computations scaling with $\nb$ (small) and not $N_h$ (large).
An affine operator structure, meaning a factorization of parameter dependence as in Eq.~\eqref{eq:affine}, is needed to achieve the desired online efficiency because size-$N_h$ operations such as $\mel{\psi_i}{H_\alpha}{\psi_j}$ are independent of $\params$ and  only need to be calculated once in the offline stage.
If the problem is non-affine, then the  strategy is to apply a so-called hyperreduction approach, which leads to an approximate affine form~\cite{Quarteroni:218966}.
A nuclear scattering example that treats a non-affine Hamiltonian is given in \textcite{Odell:2023cun}.

\subsection{Variational and Galerkin formulations}\label{subsec:other_formulations}

More generally, an RBM can be formulated in terms of
a functional that is stationary at the desired solution (variational approach) or via a weak form arising from multiplying the underlying equations and boundary conditions by arbitrary test functions and integrating over the relevant domains (Galerkin approach)~\cite{hesthaven2015certified,zienkiewicz2013finite,Brenner:2008}.
To date, EC in nuclear physics has most often been implemented with a variational formulation, which for bound energy eigenstates is familiar from introductory physics.
Less familiar, but well established, are various variational approaches to quantum scattering, where
each approach leads to a different emulator, see Sec.~\ref{sec:scattering}.

The Galerkin approach starts with the schematic form
\begin{equation}
    \mel{\zeta}{H(\params)-E(\params)}{\psi} = 0, \quad \forall\bra{\zeta} ,
   \label{eq:eigen_Galerkin}
\end{equation}
with arbitrary test functions $\ket{\zeta}$.
The reduced dimensionality for Galerkin RBM formulations enforces orthogonality with a restricted set of $\nb$ test functions,
\begin{equation}
    \ \mel*{\zeta_i}{H(\params)-\wt E(\params)}{\wt\psi} = 0, 
    \quad i = 1,\cdots,n_b .
    \label{eq:Galerkin_ortho}
\end{equation}
If the test functions are chosen to be the trial basis functions, $\bra{\zeta_i} = \bra{\psi_i}$, then this is called Bubnov-Galerkin or Ritz-Galerkin (or just Galerkin). If a different basis of test functions is used, this is called Petrov-Galerkin.
For eigenvalue problems with Hamiltonians that are bounded from below, the Ritz-Galerkin procedure yields the same equations as the variational approach.
The Petrov-Galerkin option means that the Galerkin procedure is more general.

For boundary-value partial or ordinary differential equations, there are general variational and Galerkin RBM formulations.
A projection-based emulator seeks the solution $\psi$ to
\begin{equation}
    D(\psi;\params) = 0 \mbox{ in } \Omega;\ B(\psi;\params) = 0 \mbox{ on } \Gamma , \label{eq:diffeqs}
\end{equation}
where $D$ and $B$ are operators in the domain $\Omega$ and its boundary $\Gamma$, respectively. 
There are many good references on Galerkin methods, see e.g.,  \textcite{zienkiewicz2013finite,Brenner:2008}.
The canonical example of a Poisson equation with Neumann boundary conditions is worked out in \textcite{Melendez:2022kid}.

The same RBM ingredients as for the eigenvalue problem apply here, with an integral formulation using a stationary functional such as an action $S[\psi]$,
with $\delta S = 0$ yielding Eq.~\eqref{eq:diffeqs}, or 
starting with
\begin{equation} \label{eq:galerkin_weak}
    \int_\Omega d\Omega\, \zeta\, D(\psi) + \int_\Gamma d\Gamma\, \overline\zeta\, B(\psi) = 0 ,
\end{equation}
and integrating by parts to reach a Galerkin weak formulation~\cite{zienkiewicz2013finite},
which for arbitrary test functions $\zeta$ and $\overline\zeta$ also yields Eq.~\eqref{eq:diffeqs}.
With the snapshot trial basis \eqref{eq:trial_basis}, $\delta S = 0$ can be solved for the optimal $\beta_i$ (for linear operators this is just a matrix equation).
The Galerkin formulation needs a choice for the test basis,
$\bra{\zeta} = \sum_{i=1}^{n_b}\delta\beta_i\bra{\zeta_i}$, where the $\delta\beta_i$ organize the orthogonalization conditions for each $i$:
\begin{equation}
    \delta\beta_i \Bigl[\int_\Omega d\Omega\, \zeta_i\, D(\wt\psi) + \int_\Gamma d\Gamma\, \overline\zeta_i\, B(\wt\psi) \Bigr] = 0 .
\end{equation}
(For notational simplicity we leave  partial integrations implicit here.)
Again we have Ritz-Galerkin and Petrov-Galerkin options.
For a broad set of engineering and science applications of these approaches, good starting points are~\textcite{Benner_2017aa,Benner20201,Benner2020Volume3Applications,hesthaven2015certified,Quarteroni:218966}.
Galerkin methods for quantum scattering are discussed in
Sec.~\ref{sec:scattering}, including an application to a non-affine Hamiltonian parameterization.

\begin{figure}[tbp] \centering
 \includegraphics[width=0.9\columnwidth]{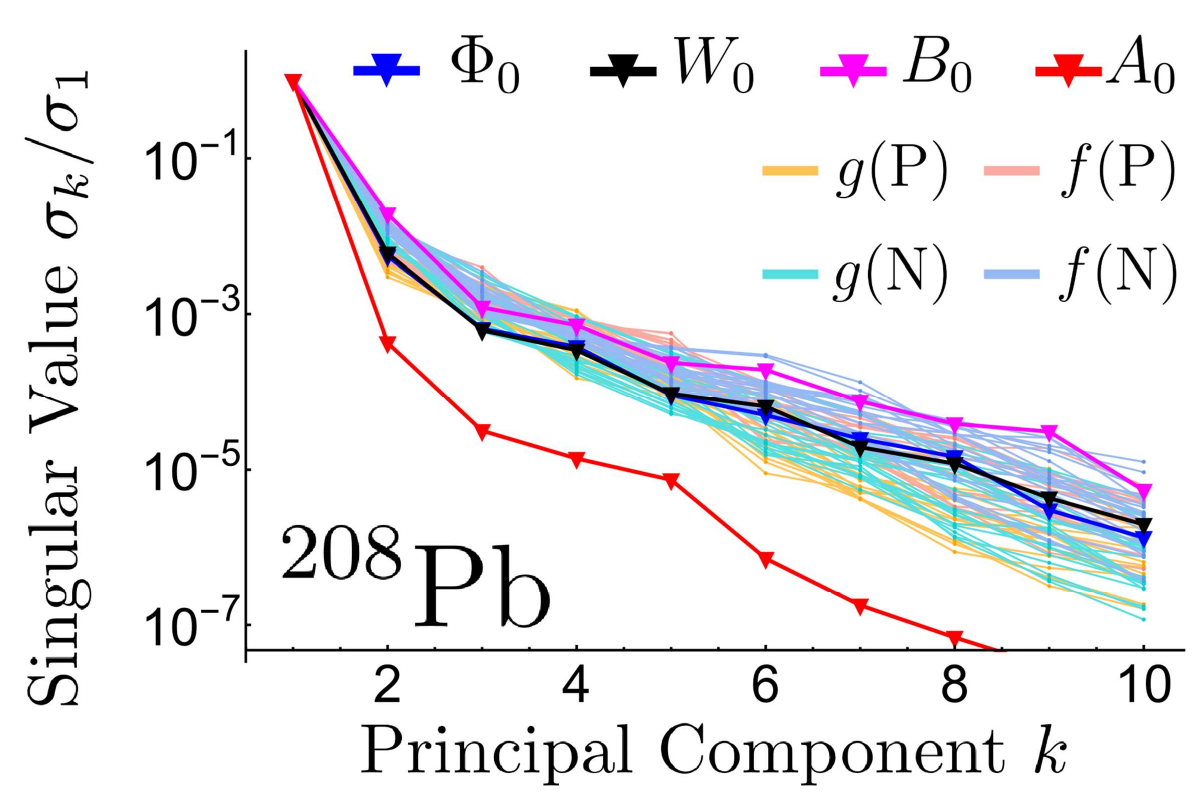}

 \caption{%
  Normalized singular values from $\nb=50$ snapshots of various functions that enter a nuclear physics energy density functional.
  The snapshots correspond to different parameter sets to be used in a Galerkin formulation of the  energy density functional, which will be solved many times with different sets for Bayesian parameter estimation.
  The rapid decrease with principal component number $k$ indicates that a small basis size will be accurate, leading in this case to speed-ups of several thousand compared to the original solver. 
  Adapted from a figure in \textcite{Giuliani:2022yna}.
  \label{fig:Singular_values_BGF}
 }
\end{figure}

A pedagogical illustration of Galerkin methods adapted to nuclear physics energy density functionals for uncertainty quantification is given in~\textcite{Giuliani:2022yna}.
Figure~\ref{fig:Singular_values_BGF} shows the singular values from snapshots of various functions that arise in the coupled nonlinear differential equations to be solved in minimizing such an energy density functional.
The efficacy of a basis obtained by POD from snapshots is implied by the rapid decrease in magnitude of the singular values, leading to high accuracy from a relatively small basis and speed-ups of order several thousand (the actual speed-up will be implementation dependent).

\subsection{Other approaches to generalized eigenvalue problems}
\label{sec:generalized}

As already emphasized,
the key to fast emulation with EC is a decomposition into one-time offline tasks and repeated, computationally efficient, online tasks. In this context, the key equations to be solved online are the low-dimensional generalized eigenvalue problem of Eq.~\eqref{eq:gen_eigen} and Fig.~\ref{fig:RBM_offline_online}(c). 
The non-orthogonality of the high-fidelity snapshots yields a non-trivial norm matrix $\wt N(\params)$ on the right-hand side of Eq.~\eqref{eq:gen_eigen}. This type of secular equation is routinely encountered in nuclear physics in the {\it discretized} version of the projected generator coordinate method~\cite{Griffin:1957zza,Brink:1968ybn,Ring:1980,Frosini:2021fjf} and in the Monte Carlo shell model~\cite{otsuka01a,shimizu12a}. The same is true for the non-orthogonal configuration interaction~\cite{thom20a} in quantum chemistry. 

Because the norm matrix is Hermitian and the EC online problem is low dimensional, the norm matrix $\wt N(\params)$ can be diagonalized to transform Eq.~\eqref{eq:gen_eigen} into a standard matrix diagonalization problem\footnote{Different numerical methods are called for when the dimension is large enough to forbid a straight diagonalization of the norm matrix; see, e.g., \cite{Frosini:2021ddm}. Notice further that the equivalence of the original and transformed secular equations is not guaranteed in the {\it continuous} version of the projected generator coordinate method~\cite{detoledopiza77a,broeckhove79a}. However, none of these two issues occur in the present context.}. Still, near-linear redundancies between the high-fidelity snapshots make the norm matrix poorly conditioned numerically. Consequently, its kernel $L_0$ must be explicitly separated from its orthogonal complement $L_{\perp}$ before transforming Eq.~\eqref{eq:gen_eigen} unitarily. In fact, 0 being an accumulation point of the eigenspectrum of the norm matrix in the limit of infinite dimension, very small non-zero eigenvalues can generate instabilities even for the finite dimensions presently under consideration. The practical remedy to this problem consists of removing eigenvectors in $L_{0}$ associated with eigenvalues smaller than a given threshold $\epsilon_{\text{th}}$ chosen such that the end results do not depend on its specific value.\footnote{This is effectively equivalent to removing the singular values below $\epsilon_{\text{th}}$ in a (truncated) POD/SVD algorithm and is standard practice in the RBM approach.}

There are several other methods used to deal with the inversion of poorly-conditioned norm matrices. Tikhonov regularization is one popular approach~\cite{Tikhonov:1943}. The simplest form of Tikhonov regularization is ridge regression or nugget regularization. In this approach, a small positive multiple of the identity is added to the norm matrix that needs to be inverted.  However, it is often not clear how to estimate the systematic bias introduced using this approach.

A new approach called the trimmed sampling algorithm was introduced in~\textcite{Hicks:2022ovs}.  Trimmed sampling uses physics-based constraints and Bayesian inference to reduce errors of the generalized eigenvalue problem.  Instead of simply regulating the norm matrix, probability distributions are sampled for the Hamiltonian and norm matrix elements, weighted by likelihood functions derived from physics-informed constraints. These physics-informed constraints include well-motivated physics principles such as positivity of the norm matrix and the smooth convergence of extremal eigenvalues with respect to variational subspace size.  The posterior distribution is determined for the Hamiltonian and norm matrix elements, and eigenvectors and observables are then sampled from that distribution. 

\begin{figure}[t]
 \includegraphics[width=0.95
\columnwidth]{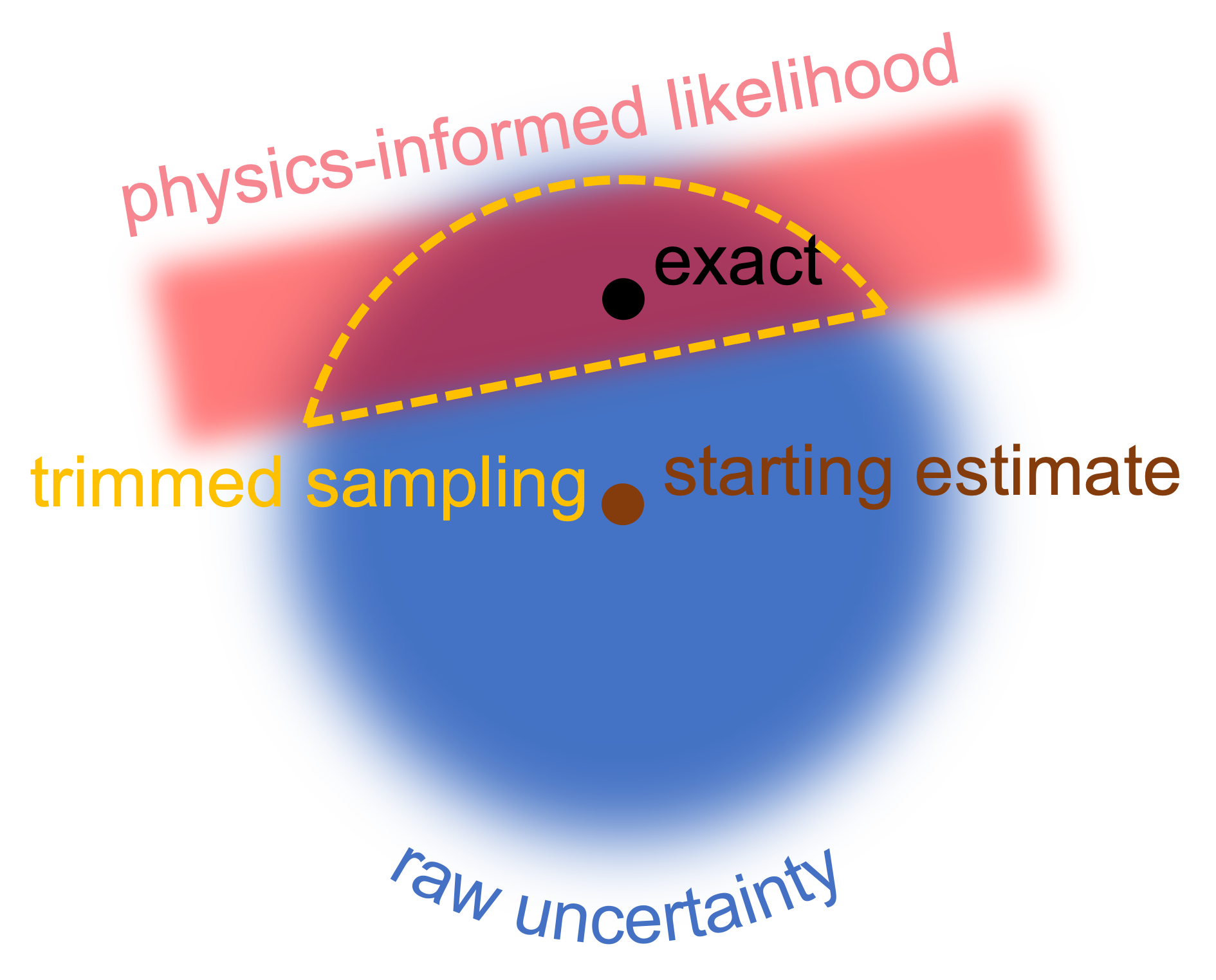}
 \caption{Schematic diagram of trimmed sampling. The raw uncertainty of some observable obtained from solving the generalized eigenvalue problem is sketched.  The raw uncertainty is centered around the starting estimate, and is used as the prior probability distribution.  The posterior probability is proportional to the product of the prior probability and the likelihood associated with the enforcement of some physics-informed constraints. The exact solution is located at a point where both the prior probability and the likelihood are not small \cite{Hicks:2022ovs}.
  \label{fig:trimmed_sampling}
 }
\end{figure}

In Fig.~\ref{fig:trimmed_sampling} a schematic diagram of trimmed sampling is displayed. The raw uncertainty of some observable obtained from solving the generalized eigenvalue problem is sketched.
The raw uncertainty centered around the starting estimate is used as the prior probability distribution. The posterior probability is proportional to the product of the prior probability and the likelihood associated with the enforcement of some physics-informed constraints. The posterior probability distribution does not give a rigorous estimate of the error. However,it can be concluded that the exact solution is located at a point where both the prior probability and the likelihood are not small~\cite{Hicks:2022ovs}.

\section{Convergence Properties of EC}\label{sec:theory}

An important and fundamental question regarding EC is how fast it converges to the exact answer as a function of the number of snapshots.  
\subsection{Bounds on the EC convergence rate}
We are interested in the rate of convergence of EC for interpolation as well as the more difficult problem of extrapolation. We start with the problem of interpolation and consider a Hamiltonian $H(\param)$ with a single control parameter, $\param$.  Let $B$ be a compact real-valued domain for $\param$ and $\ket{\psi(\param)}$ be the eigenvector of interest.  Let $d(\param,S_{n_b})$ be the norm of the residual vector when approximating $\ket{\psi(\param)}$ using EC with snapshots $S_{n_b} = \{\ket{\psi(\param_1)}, \cdots, \ket{\psi(\param_{n_b})}\}$ chosen from $B$.  We let $d_{n_b}$ denote the best possible uniform error bound for $d(\param,S_{n_b})$.  This means that we optimize the selection of snapshots for fixed ${n_b}$ such that $\max_{\param \in B}d(\param,S_{n_b})$ is minimized.  Our $d_{n_b}$ is an example of a Kolmogorov $N$-width, which is used to characterize the error and convergence of linear subspace approximations \cite{kolmogoroff1936uber,Tikhomirov_1960,pinkus2012n}.

Within its radius of convergence, a power series expansion converges exponentially fast with respect to truncation order. For example, if we truncate the power series around $\param = 0$ in Eq.~\eqref{eqn:series} at order $M$, the resulting error will be $O(|\param/z|^{M+1})$ in the limit of large $M$, where $z$ is the nearest non-analytic point.  If $\ket{\psi(\param)}$ is analytic on $B$, we can use this fundamental property of power series to derive an upper bound on EC errors when used for interpolation. 

First, we select a set of points such that all parts of $B$ lie within the radius of convergence of one of these points, which we will call anchor points.  We then take snapshots at these anchor points as well as points infinitesimally close to the anchor points.  Linear combinations of the snapshots can be used to construct derivatives and higher-order derivatives of $\ket{\psi(\param)}$ at each anchor point.  If we take ${n_b}$ snapshots, then, in the limit of large ${n_b}$, we have enough basis vectors to express the power series at each anchor point up to a truncation order that scales as $O({n_b})$.  It follows that $d_{n_b}$ is $O(x^{{n_b}})$ for some positive $x < 1$.  The generalization to compact real-valued domains in $d$ dimensions is straightforward.  We have $O(k^d)$ gradients and higher-order gradients for the multi-parameter power series in $\param_1, \cdots, \param_d$ at truncation order $k$.  For the multi-dimensional case,  $d_{n_b}$ is $O(x^{{n_b}^{1/d}})$ for some positive $x < 1$.  In the RBM literature, snapshots composed of high-fidelity solutions and their first derivatives at different anchor points have been used for partial differential equations and this is known as the Hermite subspace approach~\cite{Ito:2001}.

In general, EC extrapolation converges more slowly than interpolation.  Consider the case where the EC snapshots lie in the neighborhood of some point, but extrapolation is required beyond the radius of convergence at that point.  As illustrated in Fig.~\ref{fig:complex_plane}, we can perform secondary expansions and analytically continue past branch points in the complex plane.  We can bound the EC extrapolation error in terms of the convergence of multi-series expansions such as that shown in Eq.~\eqref{eqn:double_sum}.  These secondary expansions result in slower convergence, and the problem is most severe when there are branch points at $\param = z$ and $\param = \bar{z}$ that pinch the real axis.  The number of secondary expansions needed will scale inversely with the imaginary part of $z$.  The smaller $\Im z$, the sharper the resulting avoided level crossing.  For systems undergoing a quantum phase transition, $\Im z$ will decrease with system size and this limits the ability of EC to extrapolate across the transition point in large systems~\cite{Franzke:2023hdh}.

\begin{figure}[bt]
 \includegraphics[width=0.95
\columnwidth]{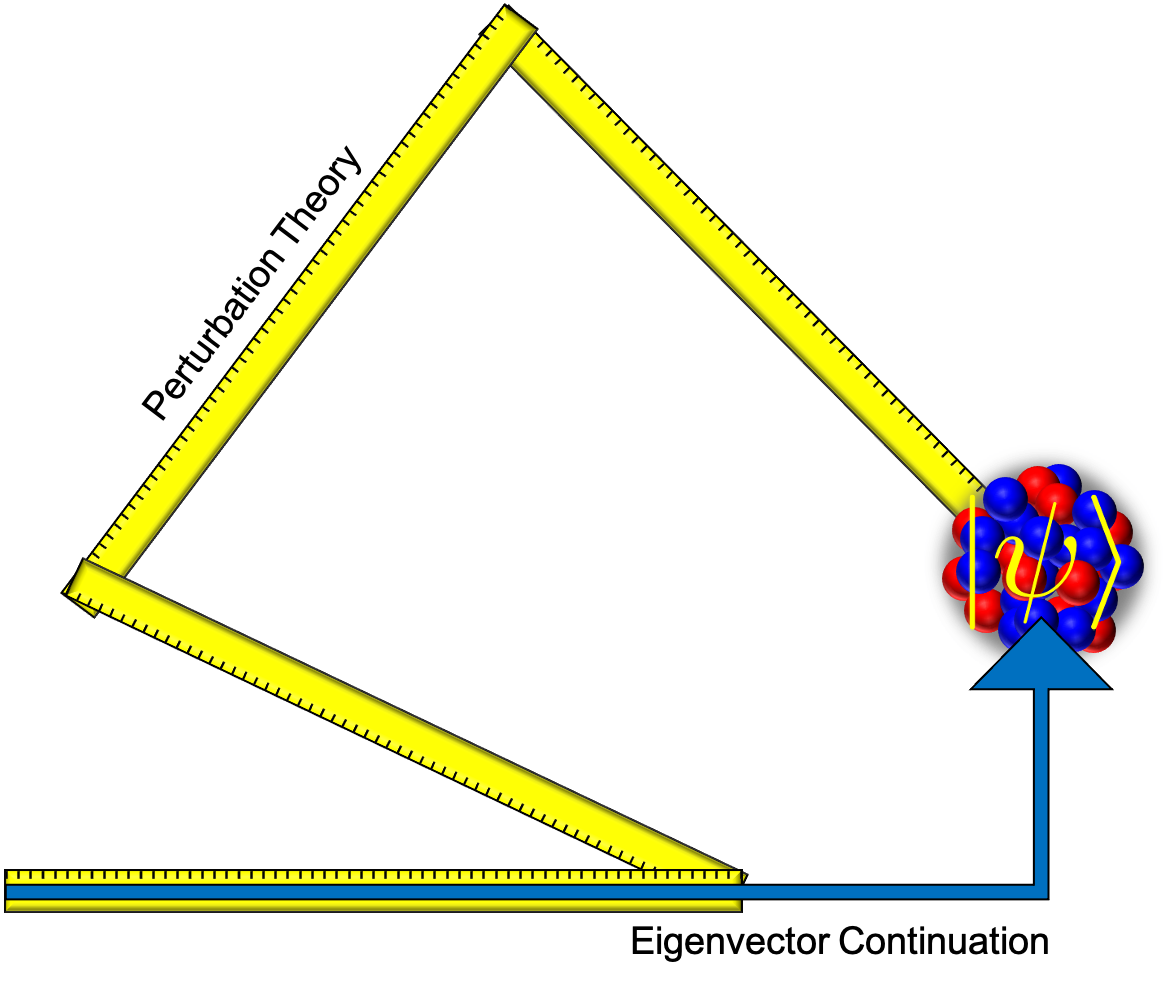}
 \caption{The convergence of perturbation theory is impacted by a phenomenon called differential folding, where cancellations occur between terms in the power series expansion for $\ket{\psi(\params)}$ \cite{Sarkar:2020mad}.  Differential folding does not occur in EC calculations, since the linear subspace is expanding along new orthogonal directions.
  \label{fig:folding}
 }
\end{figure}

The analysis described above based on power series series and perturbation theory gives an upper bound on the asymptotic error of EC for Hamiltonians $H(\params)$ that are analytic in $\params$.  However, the actual convergence rate of EC is typically faster than that of perturbation theory when selecting snapshots infinitesimally close to some anchor point.  This stems from the fact that the gradients and higher-order gradients in Eq.~\eqref{eqn:gradients} are not orthogonal to each other.  As described in~\textcite{Sarkar:2020mad} and illustrated in Fig.~\ref{fig:folding}, this results in a phenomenon called ``differential folding'' by the authors, where cancellations occur between terms in the power series expansion for $\ket{\psi(\params)}$.  No such phenomenon occurs in subspace projection methods such as EC.  As new snapshots are included, the linear subspace is expanding along directions that are orthogonal to the previous snapshots.  This produces a faster convergence for EC than perturbation theory.

\begin{figure}[t]
 \includegraphics[width=0.99
\columnwidth]{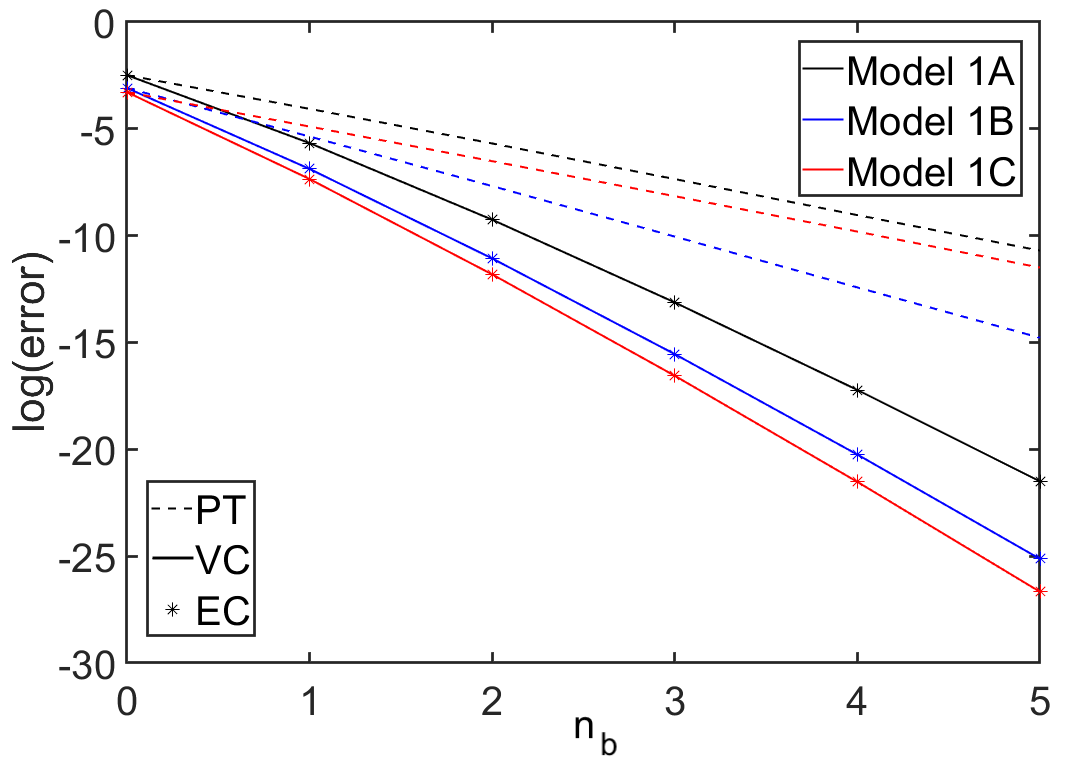}
 \caption{The logarithm of the error for the eigenstate wave function versus truncation order for perturbation theory (PT), vector continuation (VC), and eigenvector continuation (EC).  The results are for three different matrix examples described in \textcite{Sarkar:2020mad} and labelled as Model 1A, 1B, and 1C.  We see that EC and VC both converge faster than perturbation theory.
  \label{fig:PT_VC_EC}
 }
\end{figure}

The faster convergence of EC versus perturbation theory can be seen in three different matrix examples denoted as Models 1A, 1B, and 1C in \textcite{Sarkar:2020mad}.  Each of these matrix models have an affine dependence on one parameter.  In Fig.~\ref{fig:PT_VC_EC}, we show the logarithm of the error for the eigenstate wave function versus truncation order for perturbation theory (PT), vector continuation (VC), and eigenvector continuation (EC).  Vector continuation corresponds to simple projection of the exact eigenvector onto the subspace spanned by the EC snapshots.  For all three cases, we see that the VC and EC are converging significantly faster than perturbation theory.  In \textcite{Sarkar:2020mad}, it is proven that VC and EC approximations agree up to terms that scale quadratically with the error of the VC and EC approximations.

\subsection{Improved Many-Body Perturbation Theory} \label{sec:MBPT}

A natural application of EC relates to overcoming some critical limitations of many-body perturbation theory(ies) applied to nuclear systems. While more advanced (e.g., non-perturbative) expansion methods are nowadays employed to obtain accurate solutions of the nuclear many-body Schrödinger equation~\cite{Hergert:2020bxy}, many-body perturbation theories of various flavors happen to be of great use for many applications~\cite{Tichai:2020dna}. 

In this context, the generic parametric dependence of the Hamiltonian takes the simple form
\begin{equation}
H(\param) = H_0 + \param H_1 \, ,
\end{equation}
with $\param$ a complex number, knowing that the case of physical interest corresponds to $\param = 1$. Eigenstates of $H(\param)$ can be accessed via perturbation theory as a Taylor series around $\param=0$, i.e., via an expansion with respect to eigenstates of $H_0$. Eventually, an eigenstate $\ket{\Psi_n (\param)}$ of $H(\param)$ and its eigenenergy $E_n(\param)$ are approximated at perturbative order $P$ through
\begin{align}
\ket{\Psi^{(P)}_n (\param)}
 &\equiv \sum_{p=0}^{P} \param^p \ket{\Phi^{(p)}_n} \, , \\
 E^{(P)}_n(\param) &\equiv \sum_{p=0}^{P} \param^p {\cal E}^{(p)}_n \, ,
\end{align}
where the corrections $\{(\ket{\Phi^{(p)}_n}, {\cal E}^{(p)}_n); p \in  \mathbb{N}\}$ can be computed from the eigenstates of $H_0$~\cite{Shav09MBmethod}. 

The key problem relates to the fact that the sequence $\{(| \Psi^{(P)}_n(\param) \rangle , E^{(P)}_n(\param)); P \in  \mathbb{N}\}$ typically converges towards $(\ket{\Psi_n (\param)}, E_n(\param))$ when $P\rightarrow \infty$ only for $||\param||\in [0,R_c]$, where $R_c$ denotes the convergence radius. In case $R_c<1$, the problem of physical interest is inaccessible via perturbation theory.

In nuclear many-body calculations, several features can lead to $R_c<1$~\cite{Tichai:2020dna}, e.g., characteristics of internucleon interactions, choice of $H_0$, and the closed- or open-shell nature of the nucleus under study. 
While appropriately acting on the first two aspects allows one to bypass the problem in closed-shell nuclei~\cite{Tichai:2016joa}, it is much more challenging to do so in open-shell systems~\cite{Demol:2020mzd}. In this context, EC was shown to provide a systematic framework to enlarge the convergence radius via analytic continuation, i.e., the EC employing the set of P-order perturbative snapshots $\{\ket{\Psi^{(P)}_n(\param_i)}; i=1,\ldots,P+1\}$ acts as a resummation technique delivering a controlled and variational sequence of approximations to $E_n(1)$ for increasing $n_b=P+1 \in  \mathbb{N}$~\cite{Demol:2019yjt,Franzke:2021ofs}. 

\begin{figure}[tbp] \centering
 \includegraphics[width=0.95\columnwidth]{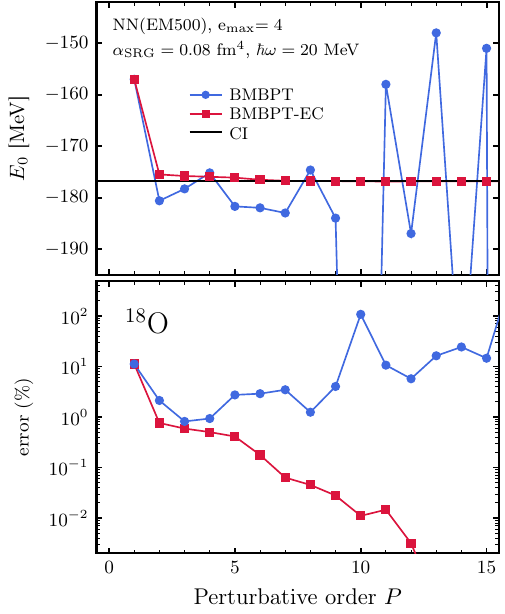}
 \caption{Ground-state energy of $^{18}$O from Bogoliubov many-body perturbation theory (BMBPT) (blue circles) and  BMBPT-based EC (red squares)  as a function of the perturbative order P against exact diagonalization (full line). The employed Hilbert space dimension is small enough for the exact diagonalization of the nuclear Hamiltonian to be accessible via configuration interaction (CI) techniques. Top panel: absolute values. Bottom panel: relative error to exact diagonalization. Adapted from~\cite{Demol:2019yjt}.
  \label{fig:PT_180}
 }
\end{figure}

Figure~\ref{fig:PT_180} demonstrates that the sequence of approximations to the ground-state energy of the open-shell $^{18}$O nucleus obtained via EC converges rapidly from above towards $E_0(1)$ even though the corresponding perturbative series diverges. Because the Hilbert space dimension employed is small enough, the results can be validated against the exact value of $E_0(1)$ obtained via the exact diagonalization of the nuclear Hamiltonian based on configuration interaction (CI) techniques. While the use of EC to resum diverging perturbative series was first dedicated to nuclear ground states~\cite{Demol:2019yjt}, it was later extended to excited eigenstates~\cite{Franzke:2021ofs}. 

Another successful application of EC is to pairing in many-body systems, see \textcite{Franzke:2023hdh} and \textcite{Baran:2022qao}.
However, the work by Franzke et al.\ also manifested a limitation of EC mentioned above, as
they found that they could not extrapolate between the normal and superfluid regimes if including snapshots from only one regime. 
That is, extrapolating between different phases of large systems will generally fail unless information on both is included 
(see also~\textcite{Sowinski:2022wdd,Brehmer2023}).

\section{Large Hamiltonian Eigensystems}\label{sec:sparse}

A powerful approach to obtaining (part of) the spectrum of the Hamiltonian
of a quantum system is the explicit diagonalization of a large (typically
sparse) Hamiltonian matrix. 
Such approaches are ideal candidates for a straightforward application of
EC as formulated by~\textcite{Frame:2017fah}, i.e., a Galerkin projection.
Since they play a crucial role in nuclear structure theory, many related applications of EC arose relatively quickly in this context.

\subsection{No-Core Shell Model Emulators}

As a first application that fueled many of the subsequent developments,
\textcite{Konig:2019adq} used a no-core shell model framework (formulated
in terms of Jacobi coordinates) to construct EC-based emulators for $A=3,4$
nucleons, i.e., the nuclei \isotope[3]{H} and \isotope[4]{He}.
In this approach, the wave function of the Hamiltonian, written as
$H = H(\params)$ with a collection of parameters $\params$, is expanded in
eigenfunctions of a harmonic-oscillator potential with a chosen
frequency.
Truncating the harmonic-oscillator basis based on a maximum number of oscillator quanta $\Nmax$
yields a (large) finite matrix that can be diagonalized. For $A=3,4$ Hamiltonians formulated in Jacobi coordinates, thereby exactly factorizing the center of mass components of the wave functions, typical matrix dimensions in the large space are $N_h \times N_h = 10^4\times 10^4$. EC for one more states can be set up directly using the coefficient vectors obtained from
Lanczos diagonalization.

The parameters $\params$ considered by \textcite{Konig:2019adq} are the
low-energy constants of the chiral effective field theory (\chieft) potential used in that work.
Overall, there are $d = 16$ individual parameters subsumed in $\vec{c}$ that
determine two- and three-nucleon interaction in the potential.
Setting up an EC emulator proceeds following the online/offline scheme described in
Sec.~\ref{sec:background} for the generic RBM workflow:
\begin{enumerate}
 \item[i)] picking a training set $\{\params_i\}_{i=1}^{\NEC}$ of \NEC\
  parameters, using space-filling Latin hypercube sampling~\cite{McKay:1979xx}
  in the $d$-dimensional parameter domain (or some subset thereof);
 \item[ii)] performing exact calculations (for the ground states of
  \isotope[3]{H} and \isotope[4]{He}, in the case of \textcite{Konig:2019adq} for
  each point in the training set, and
 \item[iii)] constructing a pair of Hamiltonian and norm matrices 
 as described in Sec.~\ref{sec:background}, for each evaluation
  of the emulator at a target parameter point $\params_*$.
\end{enumerate}
An important property of the chiral Hamiltonian is that it typically can be written as an affine
combination as introduced in Eq.~\eqref{eq:affine}, 
\begin{equation}
 H(\params) = H_0 + \sum_{k=1}^{d} \param_k H_k \,,
\end{equation}
where $H_0$ denotes the kinetic energy plus parameter-independent parts of the chiral Hamiltonian. 
This form makes it particularly efficient to evaluate the emulator at different
target points in the parameter space (the last step in the list above) because
each operator $H_k$ can be individually projected into the EC space, and this is
a one-time cost that is part of the offline emulator setup.

\begin{figure}[t]
\centering
\includegraphics[width=0.99\columnwidth]{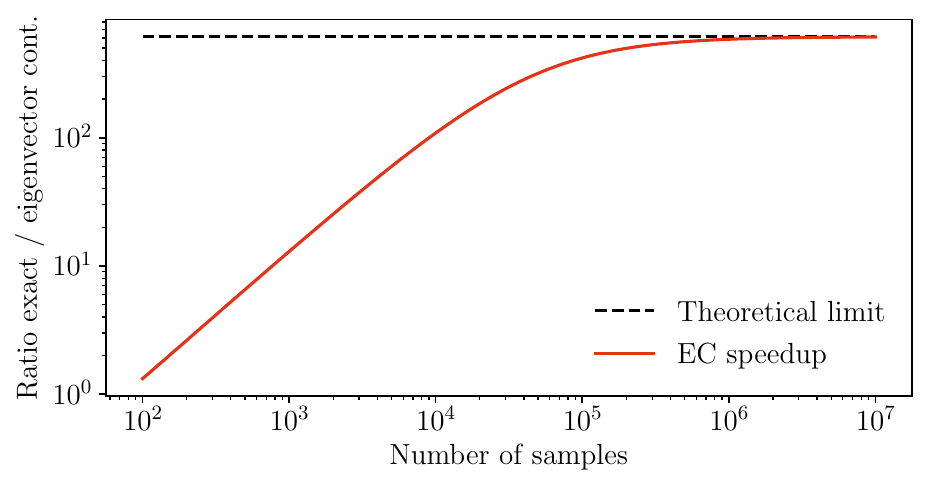}
\caption{Speed-up factor (ratio of estimated required floating-point
operations) of EC emulation compared to direct calculation as
function of the number of samples, i.e., number of calls to the
emulator (adapted from \textcite{Konig:2019adq}).
The theoretical limit indicates the maximum
speedup reached asymptotically (as the offline cost becomes amortized) in the number of samples, which is 614 for this particular example.
\label{fig:Speedup_4He_E_Nmax16-d16-n64}}
\end{figure}

\textcite{Konig:2019adq} provide a detailed analysis of the numerical performance
gain (speed-up factor) that is achieved via EC, shown in
Fig.~\ref{fig:Speedup_4He_E_Nmax16-d16-n64}
for a particular example.
While the details of that analysis are particular to the employed (Jacobi-coordinate) no-core shell-model
calculation of $^4$He, which is a light-mass nucleus of manageable computational complexity, much of the discussion applies generally to EC-based emulators with affine
parameter dependence. Most importantly, EC can be used to greatly reduce the effective dimension
of a matrix problem, and the maximum speed-up factor that follows is primarily determined by the size
of the reduced dimension compared to the original one.
The speed-up factor as a function of the number of online emulator samples shown in Fig.~\ref{fig:Speedup_4He_E_Nmax16-d16-n64} 
approaches a maximum value asymptotically because this analysis 
includes the offline cost for setting up the emulator.
For applications of RBMs to heavier-mass nuclei (discussed in Sec.~\ref{sec:cc}), that are also significantly more costly to solve for computationally using high-fidelity models, speed-up factors of the order $10^6 \cdots 10^9$ have been observed. 

\textcite{Wesolowski:2021cni} and~\textcite{Djarv:2021hcj} used EC to construct fast and accurate emulators of no-core shell-model calculations in the analysis of three-nucleon forces in \chieft. The latter work analyzed \isotope[6]{Li} in the $m$-scheme, i.e., laboratory coordinates. Thus, the dimensionality of the Hamiltonian matrix also grew dramatically. Already for $\Nmax=8$ the matrix dimension in the large space is $N_h \times N_h = 10^6 \times 10^6$, which requires significant compute efforts to be diagonalized even using the Lanczos method. \textcite{Becker:2023dqe} expressed the Hamiltonian in a symplectic symmetry-adapted no-core shell-model basis and used EC to further reduce the dimensionality of the Hamiltonian to construct accurate emulators for \isotope[12]{C}.

\subsection{Subspace-Projected Coupled Cluster}
\label{sec:cc} 

In nuclear physics and quantum chemistry one often encounters matrix representations of the many-body Schrödinger equation that are too large to permit an efficient diagonalization. For such cases, the coupled cluster (CC) method~\cite{Shav09MBmethod} can be an effective tool for approximating solutions in a space with dimensionality that is significantly reduced compared to an asymptotically exact method such as the no-core shell-model. CC is based on a similarity-transformed Hamiltonian $\overline{H}(\params) = e^{-T}H(\params)e^T$, where the cluster operator $T = T_1+T_2+\ldots+T_n+\ldots + T_A$ is the sum of $n$-particle $n$-hole excitation operators acting on an $A$-particle vacuum state $|\Phi\rangle$. In nuclear physics, $T$ is usually truncated at the singles and doubles level (CCSD), i.e., $T=T_1+T_2$, with triples excitations $T_3$ incorporated perturbatively~\cite{Hagen:2013nca}. This typically captures 99\% of the correlation energy of closed (sub-) shell systems~\cite{bartlett2007,Ekstrom:2022yea}. 

Truncating $T$ makes the similarity transformation non-unitary and $\overline{H}(\params$) non-Hermitian. In the CCSD approximation, the biorthogonal left and right eigenstates of $\overline{H}$ are obtained as solutions to the CCSD equations, which can be viewed as a set of Galerkin equations, based on 1$p$-1$h$ and 2$p$-2$h$ test functions $\ket{\Phi_i^a} \equiv a_a^{\dagger}a_i\ket{\Phi}$ and $\ket{\Phi_{ij}^{ab}} \equiv a_a^{\dagger}a_b^{\dagger}a_ja_i\ket{\Phi}$. CC calculations of atomic nuclei belong to a class of \textit{ab initio} methods that scale polynomially with system size. Still, high-fidelity and state-of-the-art calculations beyond the lightest-mass nuclei require significant high-performance computing resources. \textcite{Ekstrom:2019lss} extended EC to non-Hermitian Hamiltonian matrices and the CC method. This has paved the way for fast and accurate emulation of properties of atomic nuclei and sophisticated computational statistics analyses.

Subspace-projected CC (SPCC) is an RBM using snapshots of the bivariational left and right CC states obtained at $\nb$ different values of the parameters $\params$, e.g., the coupling constants in the description of the strong interaction Hamiltonian.   The matrix elements of the $\nb \times \nb$ SPCC Hamiltonian and norm matrices have been worked out for the case of reference states built from harmonic oscillator single-particle states~\cite{Ekstrom:2019lss}. Typically, CC calculations exploit a Hartree-Fock reference state. This embeds a dependence on $\params$ in the basis states which makes the evaluation of the matrix elements of the SPCC Hamiltonian and the norm matrices more cumbersome, but can be done using, e.g., a generalized Wick's theorem. The use of SPCC with more complex reference states are currently being explored. Inspired by the success of SPCC, an RBM of angular-momentum projected Hartee-Fock was recently applied to emulate Hartree-Fock calculations of excited states in axially deformed nuclei~\cite{Ekstrom:2023nhc}.

\begin{figure}[tbp] \centering
 \includegraphics[width=0.95\columnwidth]{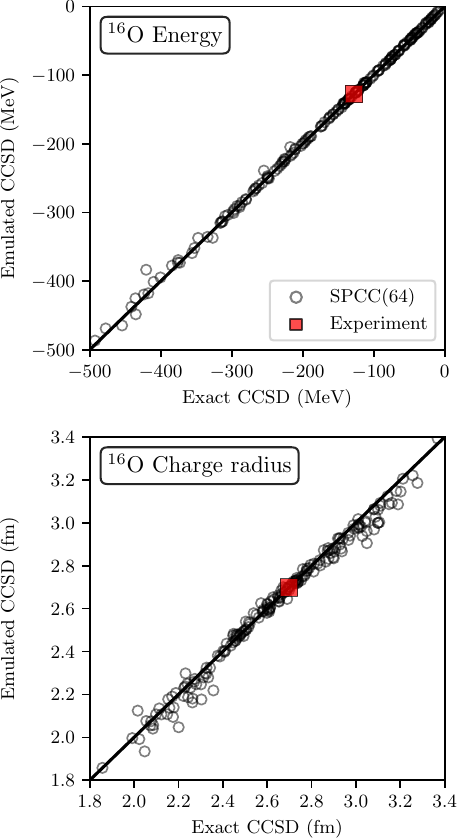}
 \caption{%
    Comparison of SPCC, based on $\nb=64$ snapshots, and exact CCSD  calculations for the ground-state energy (top panel) and charge radius (bottom panel) across a wide range of values. Adapted from~\cite{Ekstrom:2019lss}
  \label{fig:spcc}
 }
\end{figure}

The first application~\cite{Ekstrom:2019lss} of SPCC to an atomic nucleus, $^{16}$O, demonstrated that $\nb \approx 50$ CCSD snapshots are sufficient to accurately emulate, i.e., with sub-percent precision, realistic predictions of the energy and charge radius of the ground state in this nucleus as a function of $\params$. Here, $\params$ denote the 16 low-energy constants of a nuclear interaction description at next-to-next-to-leading order in \chieft. As it turns out, the accuracy of the SPCC emulator is remarkable even when using few snapshots in a very wide range of values for $\params$. Indeed, $n_b=64$ snapshots $\params_i$ in a Latin hypercube design covering an extremely dispersed set of predictions for the energy and radius in $^{16}$O is sufficient to obtain $\sim$97\% accuracy compared to exact CCSD predictions (see Fig.~\ref{fig:spcc}). Narrowing the set of snapshots to a physically motivated parameter domain increases the accuracy significantly while using even fewer snapshots. SPCC emulators are typically also very fast and the bulk properties of $^{16}$O could be sampled for one million values of $\params$ in one hour on a standard laptop, while an equivalent set of exact CCSD calculations would require 20 years of single-node compute time, i.e., an observed speed-up factor of $10^5$.

Since the first application, SPCC has been extended to emulate the properties of ground- and excited states in heavier-mass  nuclei~\cite{Hu:2021trw,Kondo2023} and infinite nuclear matter~\cite{Jiang:2022tzf} at different levels of fidelity up to perturbative triples excitations. The very low computational cost of the SPCC method, with observed speed-up factors of $10^9$, has thus enabled a wide range of exciting computational statistics analyses expounding how nuclear properties are linked to effective field theory descriptions of the strong interaction.

The CC method follows a bivariational principle which renders the SPCC Hamiltonian non-Hermitian. This may lead to difficulties identifying the target state in the spectrum as it is not guaranteed to be the lowest. Indeed, in heavier-mass nuclei and nuclear matter, where level densities are higher, the target state can sometimes appear as an excited state. The interpretation of the SPCC spectrum under these conditions remains an open challenge. Using the bivariational principle, ~\textcite{Jiang:2022oba} introduced a method called ``small-batch voting'' to detect the target state in such scenarios. Much remains to be discovered regarding the convergence properties of non-Hermitian EC and the advantages of RBMs applied to the CC method.

\subsection{Phenomenological shell model}

While emulating low-energy constants stemming from \chieft is probably one of the most
relevant scenarios in \abinitio nuclear structure theory, EC provides
opportunities to enhance phenomenological approaches as well.
One such application is given by the work of \textcite{Yoshida:2021jbl}, who
apply EC in connection with the nuclear shell model.
In a shell-model calculation, the Hamiltonian is typically split into one- and
two-body terms, $H = H^{(1)} + H^{(2)}$, with natural extensions to include
higher-body terms.
$H^{(1)}$ models the effective mean field that generates nuclear orbitals, while
$H^{(2)}$ describes interaction among valence nucleons.
Input parameters for a shell-model calculation are single-particle energies, determining the diagonal part of $H^{(1)}$, and two-body matrix elements
that parametrize the interaction in $H^{(2)}$.
Both types of parameters need to be fitted to experimental data within the
regime of nuclei that one wishes to describe with a particular model (e.g., $sd$
shell nuclei).

As a typical starting point, \textcite{Yoshida:2021jbl} consider the USDB
interaction for $sd$ shell nuclei~\cite{Brown:2006gx}, which spans overall a
66-dimensional parameter space of 3 single-particle energies and 63 two-body matrix elements, collected into a vector
$\params$.
While each individual diagonalization within the valence space can be quite
cheap, the large number of parameters implies that there is significant
potential for speeding up the fitting process via EC emulation.
Setting up such an emulator follows the standard EC procedure based on training
points $\params_i$ with $i=1,\cdots,\NEC$ together with the
Hamiltonian
\begin{spliteq}
 \wt{H}_{ij}
 &= \bra{\psi(\params_i)} H(\params_{\target}) \ket{\psi(\params_j)} \\
 &= \sum_k h_k^{(1)} \times \overline{\text{OBTD}}_k
 + \sum_k V_k^{(2)} \times \overline{\text{TBTD}}_k \,,
\label{eq:H-EC-shell-model}
\end{spliteq}
where $h_k^{(1)}$ and $V_k^{(2)}$ are, respectively, the single-particle energies and two-body matrix-elements that
multiply one- and two-body transition densities ($\overline{\text{OBTD}}$ and $\overline{\text{TBTD}}$), and $\ket{\psi(\params_i)}$
denotes a particular shell-model wave function obtained for parameters $\params_i$.
Importantly, Eq.~\eqref{eq:H-EC-shell-model} has an affine structure that
enables the previously mentioned online/offline decomposition.

A number of benchmark scenarios are considered, varying the number of training
points $\nb$ between 50 and 250, and also the number of states (lowest part of
the spectrum starting with the ground state) per training parameter set, between
1 and 5.
Overall, for a selection of nuclei such as \isotope[28]{Si} and \isotope[24]{Mg}
relative emulator errors of the order between less than one percent up to a few
percent are observed.
A Monte Carlo sampling technique is proposed to assign emulator uncertainties
for individual evaluations.

In line with other work such as \textcite{Konig:2019adq},
\textcite{Yoshida:2021jbl} find that emulated wave functions generally show
larger emulation discrepancies than binding energies, leading to a larger spread
for emulator evaluations of operators such as magnetic dipole moments and
quadrupole moments.
To improve the emulator accuracy and avoid problems in correctly describing
such observables, \textcite{Yoshida:2021jbl} suggest to use the shell-model
emulator as a preprocessor to generate optimized
initial states for a subsequent exact Lanczos diagonalization.
More generally, there will be challenges in applying EC beyond the $sd$ shell (e.g., to the $pf$ shell), where there are many more parameters and the time to generate low-lying states in the offline training phase will be greater.
This is where the experience from the RBM community in reduced-order sampling (e.g., using SVD methods, see Sec.~\ref{sec:MOR}) could be profitably carried over to nuclear problems.

\section{Examples of Extensions}

In the following subsections we introduce three extensions of the basic EC method.

\subsection{Emulators for Quantum Scattering} \label{sec:scattering}

Uncertainty quantification will often require calculations of scattering observables with many different Hamiltonian parameterizations. 
Examples in nuclear physics include the calibration of \chieft interactions and of phenomenological optical potentials.
This has motivated the extension of model-driven emulators to the quantum mechanical two-body scattering problem and beyond. 
Of particular importance for nuclear applications is the ongoing development of three-body scattering emulators.

Quantum scattering is not an eigenvalue problem, but the same principles that make EC effective for bound states carry over to scattering.
In the time-independent formulation of scattering, we still start with the strong form of the Schr\"odinger equation, $H(\params)\ket{\psi(\params)} = E\ket{\psi(\params)}$, 
but now $E$ is specified rather than determined (although it can also be treated as a parameter of the emulator).
The freedom to formulate the Schr\"odinger equation for scattering in different ways, including homogeneous or inhomogeneous differential equations for scattering wave functions as well as integral equations for scattering matrices, leads to many possible emulators.
In addition there is the freedom to choose trial and test bases for Galerkin projection (see Sec.~\ref{sec:MOR}).
As such, beyond the intrinsic use of scattering emulators, the scattering problem is a prototype for multiple approaches to model reduction in other settings.

\begin{figure}[tbh]
  \includegraphics{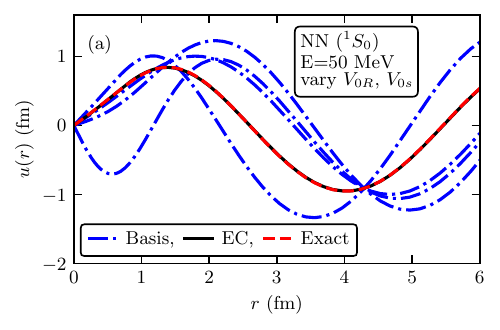}
  \includegraphics{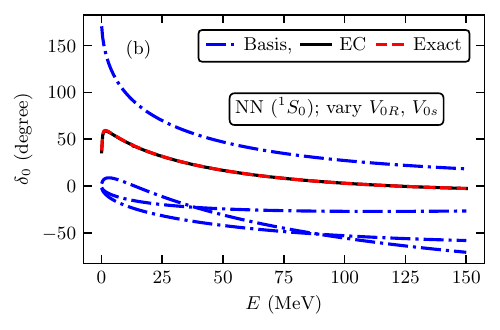}
    \caption{(a) Scattering wave functions for a model nucleon-nucleon potential at a fixed energy~\cite{Furnstahl:2020abp}.
    The dot-dashed curves are for four choices of $\params_i = \{V_{0R},V_{0s}\}$ that comprise the  trial basis, the dashed curve is for the target values, 
    and the solid curve is the prediction using the KVP emulator. 
        The curves have a common crossing point at the value of $r$ where the second term in Eq.~\eqref{eq:KVP_constraint} is zero.
    (b) Scattering phase shifts for the same parameter sets and the emulator prediction.} 
  \label{fig:NN2dim1S0_NB_4_E_50_WF}
\end{figure}

There are numerous variational formulations of scattering, such as those due to Kohn, Schwinger, and Newton~\cite{newton2002scattering}.
Variational here means that there is a stationary functional, but in most cases this does not imply that the result is an upper bound, unlike the case with bound states.
For each of these variational formulations there is a corresponding RBM emulator.

The first implementation of a quantum scattering emulator~\cite{Furnstahl:2020abp} used the
Kohn Variational Principle (KVP) for partial wave scattering~\cite{Kohn:1948col}. For two-body scattering in a single channel with angular momentum $l$ at on-shell energy $E=q^2/2\mu$, the KVP functional takes the form
\begin{equation}
    \mathcal{K}[\wt\psi] = \wt K_E + \mel*{\wt\psi}{H - E}{\wt\psi} .
    \label{eq:KVPfunctional}
\end{equation}
In Eq.~\eqref{eq:KVPfunctional} the trial scattering wave function $\ket{\wt\psi}$ in position space is constrained to satisfy the asymptotic normalization condition
\begin{equation}
    \wt\psi_l(r) \underset{r\rightarrow\infty}{\longrightarrow} j_l(qr) + n_l(qr)\tan\delta_l,
    \label{eq:KVP_constraint}
\end{equation}
and
\begin{equation}
    \wt K_E = -\frac{\tan\delta_l}{2\mu q}
    \label{eq:Kmatrix}
\end{equation}
is the on-shell $K$-matrix corresponding to phase shift $\delta_l(E)$.
This functional is stationary about the exact solution $\psi$ such that $\mathcal{K}[\psi+\delta\psi] = K_E + \mathcal{O}(\delta K)^2$.

An EC/RBM emulator for the KVP uses a snapshot trial basis as in Eq.~\eqref{eq:trial_basis}, where each basis wave function satisfies Eq.~\eqref{eq:KVP_constraint} and the overall constraint for the trial wave function, which requires $\sum_{i=1}^{n_b} \beta_i = 1$, is enforced by a Lagrange multiplier. 
Varying the KVP functional with this constraint yields a low-dimensional $(n_b\times n_b)$ linear matrix problem.  If the Hamiltonian is affine in the parameters, all of the relevant matrix elements can be pre-computed in the offline stage as in Fig.~\ref{fig:RBM_offline_online}.
An example of this emulator from \textcite{Furnstahl:2020abp} is shown in Fig.~\ref{fig:NN2dim1S0_NB_4_E_50_WF} for a model nucleon-nucleon potential with two parameters (the strengths of two Gaussians).
The snapshot wave functions for four randomly chosen sets of $\params_i$ are shown in panel \ref{fig:NN2dim1S0_NB_4_E_50_WF}(a) while the corresponding phase shifts are shown in panel \ref{fig:NN2dim1S0_NB_4_E_50_WF}(b). 
Despite no indication from the figure that this is a good basis, the emulator is fast and accurate through the full range of energies.

The KVP is sometimes itself used as a high-fidelity solution method, where it is is well-known to be plagued with numerical issues known as Kohn anomalies. 
These can be mitigated for emulators by a more general formulation than Eqs.~\eqref{eq:KVPfunctional}--\eqref{eq:Kmatrix} that uses multiple scattering matrices (rather than just the K-matrix); see \textcite{Drischler:2021qoy,Drischler:2022ipa} for details.
This approach has been extended to coupled channels and to momentum space in~\textcite{Garcia:2023slj}, with successful tests of the full range of two-body scattering observables using a state-of-the-art \chieft Hamiltonian with 25 parameters (up to six in each partial wave channel, which are emulated independently). 
Speed-ups of two orders of magnitude over high-fidelity calculations were found even when using basis sizes large enough to achieve a mean relative emulator error of order $10^{-10}$ over a wide region in parameter space (in practice this means $n_b$ is equal to twice the number of parameters in a given channel).

Another form of the KVP-type emulator avoids using a Lagrange multiplier to constrain the normalization of basis wave functions by introducing a trial basis only for the second (scattering) term of Eq.~\eqref{eq:KVP_constraint} rather than the full wave function.
The free wave function (first term in Eq.~\eqref{eq:KVP_constraint}) fixes the normalization. 
The Schwinger and Newton emulators use alternative variational principles, with the latter having a trial basis of K-matrices rather than wave function~\cite{Melendez:2021lyq}; it is applied to the calibration of \chieft parameters in~\textcite{Svensson:2023twt}.

Each of these variational formulations has a Galerkin counterpart, so we can use Galerkin projection as an alternative path to constructing the emulators. This is worked out for each of the Kohn, Schwinger, and Newton emulators in \textcite{Drischler:2022ipa}.
This also means we can directly formulate scattering emulators that do not have an obvious variational counterpart.
With the normalization fixed at the origin ($r=0$) by a free solution value and first derivative, the snapshot basis of scattering terms can be used in a Galerkin projection of Eq.~\eqref{eq:eigen_Galerkin}. 
An application of this emulator to calibrate phenomenological optical potentials in \textcite{Odell:2023cun}, implemented with ROSE software from the BAND project~\cite{bandframework}, uses proper orthogonal decomposition (see Sec.~\ref{subsec:workflow}) to optimize the basis and 
demonstrates a method
to handle the non-affine parameters of the potential.
Yet another formulation builds on R-matrix theory with successful applications to fusion observables~\cite{Bai:2021xok,Bai:2022hjg}.
The frontier for scattering emulators is for three-body problems. 
A proof-of-principle demonstration using the KVP for three bosons was given in \textcite{Zhang:2021jmi} and tests of realistic nuclear scattering are in progress.

\subsection{Finite Volume Dependence and Resonances}\label{sec:volume}

Another extension of EC, developed by
\textcite{Yapa:2022nnv}, is concerned with extrapolating or interpolating
the volume dependence of energy levels in finite periodic boxes, with the
particular application of studying resonance properties via finite-volume (FV)
simulations~\cite{Wiese:1988qy,Luscher:1991cf,Rummukainen:1995vs,Klos:2018sen}.
In this scenario, not only does the Hamiltonian $H = H(L)$ depend explicitly
on the size of a cubic box $L$ (via the periodic extension of the interaction
part), but, since eigenstates of $H(L)$ have to satisfy the periodic boundary
condition, they also carry an \emph{implicit} dependence on $L$.
Specifically, states defined in boxes with different $L$ are vectors in
distinct Hilbert spaces, which makes it \textit{a priori} difficult to give
a well-defined meaning to matrices
\begin{subalign}[eq:H-N-naive]
 \wt{H}_{ij}(L_{\target}) &= \braket{\psi_{L_i}|H({L_{\target}})|\psi_{L_j}} \,, \\
 \wt{N}_{ij} &= \braket{\psi_{L_i}|\psi_{L_j}}
\end{subalign}
that appear in a standard EC setup, i.e., a Galerkin projection with snapshots defined at with training points $L_i$ and target
volume $L_{\target}$.

\textcite{Yapa:2022nnv} overcome this issue by defining a space
\begin{equation}
 \mathcal{H} = \bigcup_{\{L>0\}} \mathcal{H}_L \,.
\end{equation}
as a union of Hilbert spaces $\mathcal{H}_L$ that contain periodic states with
fixed period $L$.
This set $\mathcal{H}$ is not a vector space with the standard pointwise
addition of functions (assuming that the $\mathcal{H}_L$ are simple function
spaces), but it can be made into one by defining appropriate operations that
combine functions with different periods.

\textcite{Yapa:2022nnv} accomplish this by applying
dilatations (transformations involving stretching and rescaling) that map
states into a common space prior to applying the standard operations within that
space and show that, when this procedure is applied to a truncated bases of
periodic functions (such as simple plane waves or a discrete variable
representation as used by \cite{Klos:2018sen} to study few-body
systems in finite volume), ultimately the outcome is equivalent to simply
operating within the $\mathbbm{R}^n$ space of coefficient vectors (where
$n$ denotes the dimension of the finite space).
We note that in the broader RBM context problems such as the volume dependence
discussed here have been treated by mapping the physical domain to a fixed
reference domain and formulating an equivalent problem on this reference 
domain~\cite{Rozza:2005,Rozza:2007aa}.

\begin{figure}[tbp] \centering
 \includegraphics[width=0.95\columnwidth]{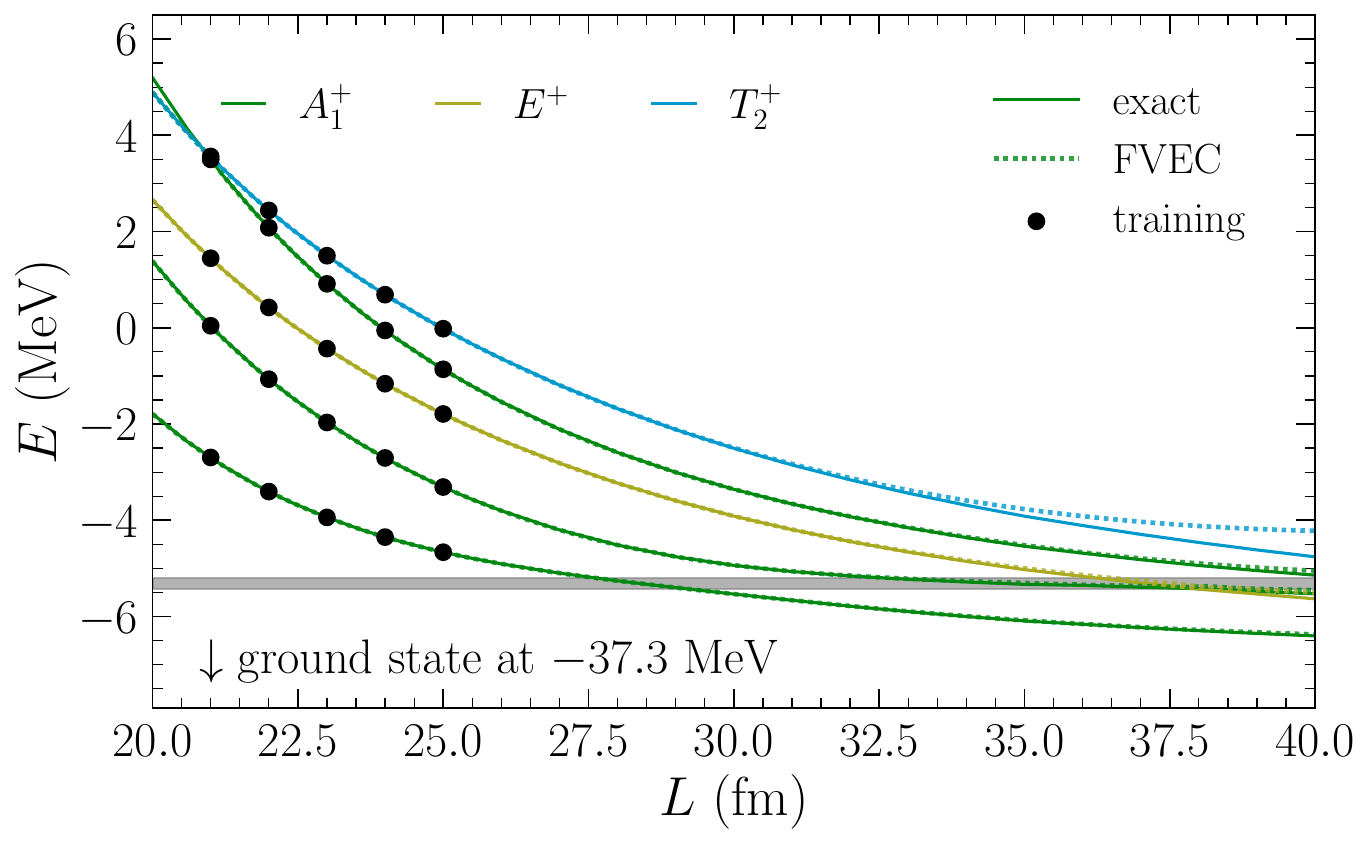}
 \caption{%
  Positive-parity finite-volume energy spectrum of three bosons exhibiting
  a resonance state.  Solid lines show the exact states
  calculated in a certain basis of the discrete variable
representation (see \textcite{Yapa:2022nnv} for details),
  whereas dashed lines indicate FVEC results obtained based on training data
  at five different box sizes (solid circles), using the 8 lowest state in
  the spectrum at each volume (including the ground state which is not
  shown in the figure).
  \label{fig:FVEC-3b-Blandon-1-Pp}
 }
\end{figure}

As mentioned at the outset of this section, studying resonances in FV was a
primary motivation for the development of FVEC.
Figure~\ref{fig:FVEC-3b-Blandon-1-Pp}, taken from \textcite{Yapa:2022nnv},
demonstrates this application with the example of a three-boson resonance,
generated by a sum of attractive and repulsive Gaussian
potentials~\cite{Blandon:2007aa}.
For almost the entire range of volumes shown in the figure, FVEC produces
results that in the resolution of the plot are virtually indistinguishable
from exact calculations, and the avoided level crossing around $L \sim 28$ fm,
indicating the resonance, is well reproduced.

Resonances are a fascinating phenomenon found in many areas of physics, closely
related to the study of open quantum systems.
Numerically studying their properties is notoriously challenging because
accommodating states that decay with a finite lifetime requires either a
time-dependent treatment, or special ``tricks'' to describe them within a
time-independent framework.
Enclosing the system in a finite volume and looking for avoided level crossing
in the volume-dependent energy spectrum is an elegant way of identifying
resonances, but this approach is geared primarily towards few-body systems.
In formal scattering theory, decaying resonances are generally associated with
poles of the scattering matrix (``$S$-matrix'') at complex energies
$E = E_R - i\Gamma/2$, located in the fourth quadrant of the complex plane.
The real part $E_R$ denotes the resonance position, while the width $\Gamma>0$
is related to the inverse of the lifetime.

While ordinary Hermitian quantum mechanics can only describe either bound states
(real $E<0$) or scattering (real $E>0$), different options to achieve
non-Hermitian extensions have been developed in order to allow for complex
energy eigenvalues.
\textcite{Yapa:2023xyf} developed an extension of EC that uses the so-called
(uniform) complex scaling technique to describe resonances, and in particular
their trajectories in the complex plane under variation of the Hamiltonian,
written as $H = H(\param)$.
Complex scaling is based on rotating radial coordinates according to $r \to r
e^{i\phi}$ with an angle $\phi > 0$, or, equivalently~\cite{Afnan:1991kb},
the conjugate momentum variable $q$ according to $q \to q e^{{-}i \phi}$.
Along this rotated contour, resonance wave functions behave effectively like
bound states, and the complex-scaled (rotated) Hamiltonian allows for complex
energy eigenvalues.
An important aspect of complex scaling (as well as other methods that enable
the description of resonance in time-independent quantum mechanics) is that
inner products of complex-scaled states do \emph{not} involve complex
conjugation of the ``bra'' state.
\textcite{Yapa:2023xyf} show that eigenvector continuation for resonance states
can be implemented by defining the Hamiltonian and norm matrix elements in terms
of the so-called ``c-product''~\cite{Moiseyev:1978aa,Moiseyev:2011}, which
for eigenstates $\ket{\psi_1}$ and $\ket{\psi_2}$ with equal angular-momentum
quantum numbers is given by
\begin{equation}
 \braket{\psi_1 | \psi_2} = \int \dd r \, \psi_1(r)\psi_2(r) \,.
\label{eq:CP-vecr}
\end{equation}

\begin{figure}[htb]
\centering
\includegraphics[width=0.95\columnwidth]{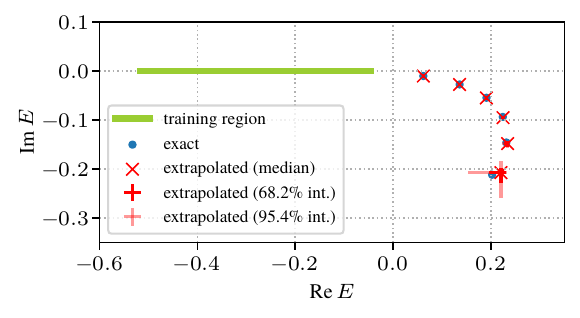}
\caption{Bound-state-to-resonance extrapolation performed with CA-EC for a
 two-body system supporting an S-wave resonance for certain values of the
 parameter $c$.
 Five training points were randomly drawn from the region $c\in (0.9,1.3)$ per
 dataset, leading to bound states within the shaded line along the negative real
 axis.
 Multiple samples of such five points were used to obtain extrapolations with
 uncertainty estimates, see \cite{Yapa:2023xyf} for details.
 \label{fig:CAEC}
}
\end{figure}

Standard EC works well in this way for extrapolating (or interpolating) the
trajectory of a resonance state as it moves in the fourth quadrant as a function
of $\param$.
While this is relevant for example for constructing EC-based emulators for
resonance properties, \textcite{Yapa:2023xyf} are furthermore interested in the
case where EC is trained within a regime of $\param$ where the states is actually
\emph{bound}, and then extrapolate from there into the resonance domain.
The key result of this work is that while using the c-product alone is not
sufficient to achieve this, an extension of EC that includes for each training
bound-state also its complex conjugate (with complex scaling, bound-state
eigenvalues remain real, but the corresponding wave functions defined along the
rotated contour have nontrivial complex behavior).
This ``conjugate-augmented eigenvector continuation (CA-EC)'' is then able to
perform the desired extrapolation from bound states to resonances.
An example for this is shown in Fig.~\ref{fig:CAEC}.

While \textcite{Yapa:2023xyf} consider as proof of concept only two-body
resonances calculated with complex scaling, the authors conjecture that
generally CA-EC is expected to work for quantum systems involving more
particles, as well as in conjunction with techniques other than complex scaling.

\subsection{Quantum Monte Carlo Simulations}\label{sec:QMC}

Quantum Monte Carlo simulations are widely used for first-principles calculations of quantum many-body systems across many subfields of physics.  In cases where sign oscillations are not a problem, the computational effort usually scales as a low-order polynomial in the number of particles.  Since quantum Monte Carlo can work with vectors in extremely large linear spaces, the combination of EC with quantum Monte Carlo methods is potentially very powerful.  The application of EC with quantum Monte Carlo is discussed in \textcite{Frame:2017fah} as well as in \textcite{Frame:2019jsw}.  

In EC we perform a Galerkin projection and thus we need to compute inner products between energy eigenstates associated with different Hamiltonians.  However, computing the inner product of the different eigenstates is not straightforward using quantum Monte Carlo simulations.  We illustrate the problem with an example involving ground state wave functions.  Let $H_A$ and $H_B$ be two quantum Hamiltonians with ground state wave functions $\ket{v^0_A}$ and $\ket{v^0_B}$, respectively, and ground state energies $E^0_A$ and $E^0_B$, respectively.  Let $\ket{\phi}$ be any state that is not orthogonal to $\ket{v^0_A}$ and $\ket{v^0_B}$.  Starting with the state $\ket{\phi}$, we can obtain $\ket{v^0_A}$ by applying the Euclidean time evolution operator $e^{-H_A t}$ and taking $t$ to be large and positive.  Similarly, we can obtain $\ket{v^0_B}$ by applying the Euclidean time evolution operator $e^{-H_B t}$.  In the limit of large $t$, we have 
\begin{gather}
    e^{-H_A t} \ket{\phi} \approx e^{-E^0_A t} \braket{v^0_A|\phi} \ket{v^0_A}, \\
    e^{-H_B t} \ket{\phi} \approx e^{-E^0_B t} \braket{v^0_B|\phi} \ket{v^0_B}.
\end{gather}
The difficulty arises from the fact that $\ket{v^0_A}$ and $\ket{v^0_B}$ appear with exponential factors of $e^{-E^0_A t}$ and $e^{-E^0_B t}$, respectively.  Calculations of the magnitude of the inner product $\braket{v^0_A|v^0_B}$ are prone to large relative errors since the amplitude is dominated by factors of $e^{-E^0_A t}$ and $e^{-E^0_B t}$ for large $t$.

A technique called the floating block method was introduced in \textcite{Sarkar:2023qjn} that addresses this problem.  The floating block method is based the observation that
\begin{equation}
\lim_{t \rightarrow \infty}\frac{\braket{\phi|e^{-H_A t}e^{-H_B t}e^{-H_A t}e^{-H_B t}|\phi}}{\braket{\phi|e^{-2H_A t}e^{-2H_B t}|\phi}}=|\braket{v^0_A|v^0_B}|^2. \label{eq:FBequation}
\end{equation}
We note that the problematic exponential factors of $e^{-E^0_A t}$ and $e^{-E^0_B t}$ cancel from this ratio.  We can also calculate the complex phase of the inner product using
\begin{equation}
\lim_{t \rightarrow \infty}\frac{\braket{\phi|e^{-2H_A t}e^{-2H_B t}|\phi}}{|\braket{\phi|e^{-2H_A t}e^{-2H_B t}|\phi}|}=\frac{\braket{v^0_A|v^0_B}}{|\braket{v^0_A|v^0_B}|}. \label{phase}
\end{equation}  
Here we are using the phase convention that $\braket{v^0_A|\phi}$ and $\braket{v^0_B|\phi}$ are positive.

\begin{figure}
 \includegraphics[width=0.9
\columnwidth]{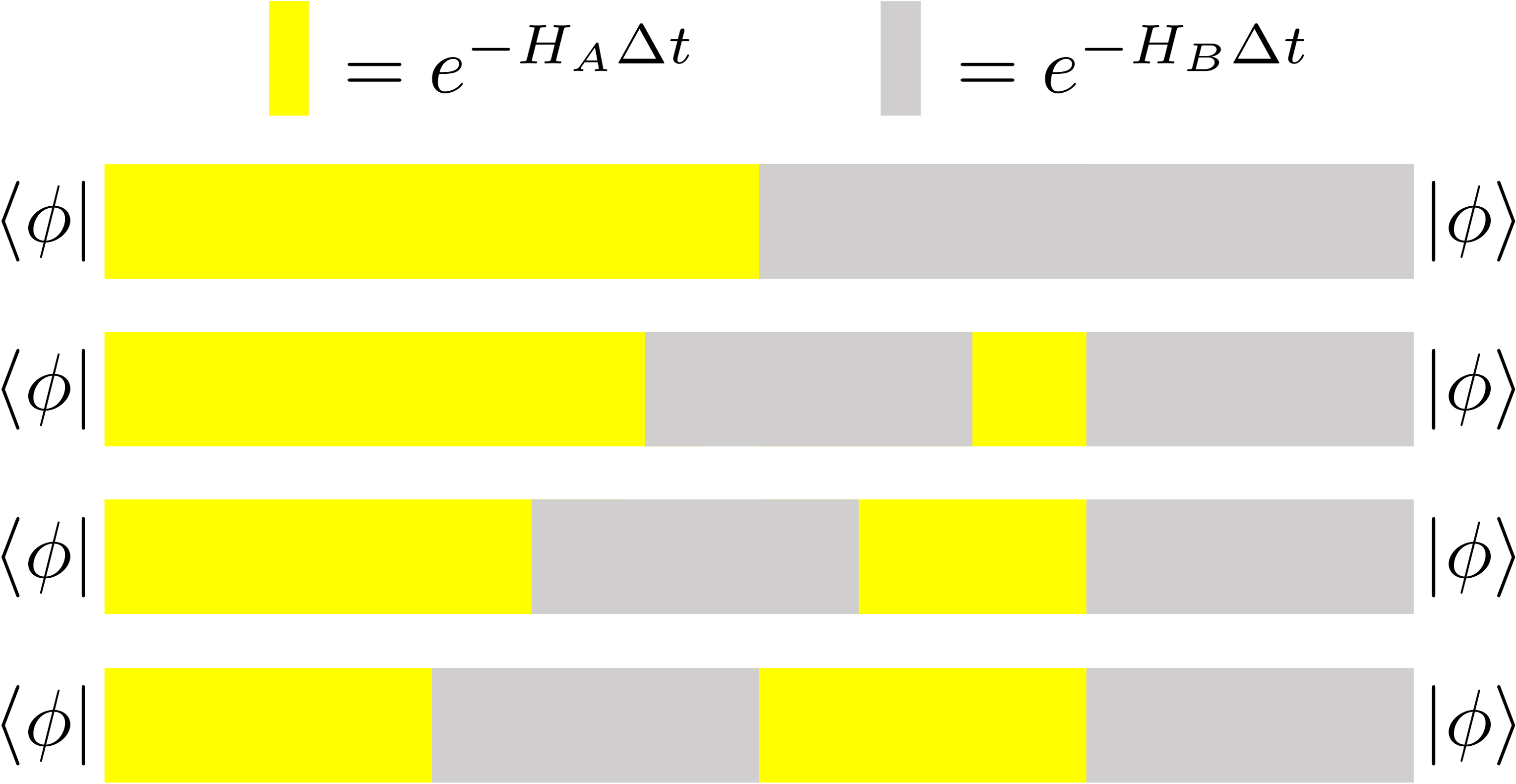}
 \caption{Schematic diagram showing the intermediate values for the Euclidean time blocks, gradually interpolating between $\braket{\phi|e^{-2H_A t}e^{-2H_B t}|\phi}$ in the denominator of Eq.~\eqref{eq:FBequation} and
 $\braket{\phi|e^{-H_A t}e^{-H_B t}e^{-H_A t}e^{-H_B t}|\phi}$ in the numerator of Eq.~\eqref{eq:FBequation} 
  \cite{Sarkar:2023qjn}. This corresponds to gradually changing the values of $t_1, t_2, t_3, t_4$ in Eq.~\eqref{eq:t1t2t3t4}. \label{fig:floating_block}
 }
\end{figure}

If we try to compute the ratio of the numerator and denominator in Eq.~\eqref{eq:FBequation} directly using Monte Carlo simulations, the result will still be noisy since the numerator and denominator are uncorrelated with each other.  In order to overcome this problem, the floating block method instead computes ratios of quantities that are strongly correlated.  Let us define $Z(t_1,t_2,t_3,t_4)$ to be the amplitude
\begin{equation}
 Z(t_1,t_2,t_3,t_4) = \braket{\phi|e^{-H_A t_1}e^{-H_B t_2}e^{-H_A t_3}e^{-H_B t_4}|\phi}. \label{eq:t1t2t3t4}
\end{equation}
In the floating block method, we compute ratios of the form
\begin{equation}
\frac{Z(t_1,t_2,t_3,t_4)}{Z(t'_1,t'_2,t'_3,t'_4)},
\end{equation}
for values $t_1,t_2,t_3,t_4$ and $t'_1,t'_2,t'_3,t'_4$ that are close to each other.  We can then form telescoping products of such ratios,
\begin{equation}
\frac{Z(t_1,t_2,t_3,t_4)}{Z(t'_1,t'_2,t'_3,t'_4)}
\frac{Z(t'_1,t'_2,t'_3,t'_4)}{Z(t''_1,t''_2,t''_3,t''_4)}
\frac{Z(t''_1,t''_2,t''_3,t''_4)}{Z(t'''_1,t'''_2,t'''_3,t'''_4)} \cdots. \label{eq:ratios}
\end{equation}
In this manner, we can calculate the ratio of the numerator and denominator in Eq.~\eqref{eq:FBequation}.  This is illustrated schematically in Fig.~\ref{fig:floating_block}.  

\begin{figure}[hbt!]
    \centering
    \includegraphics[width=9cm]{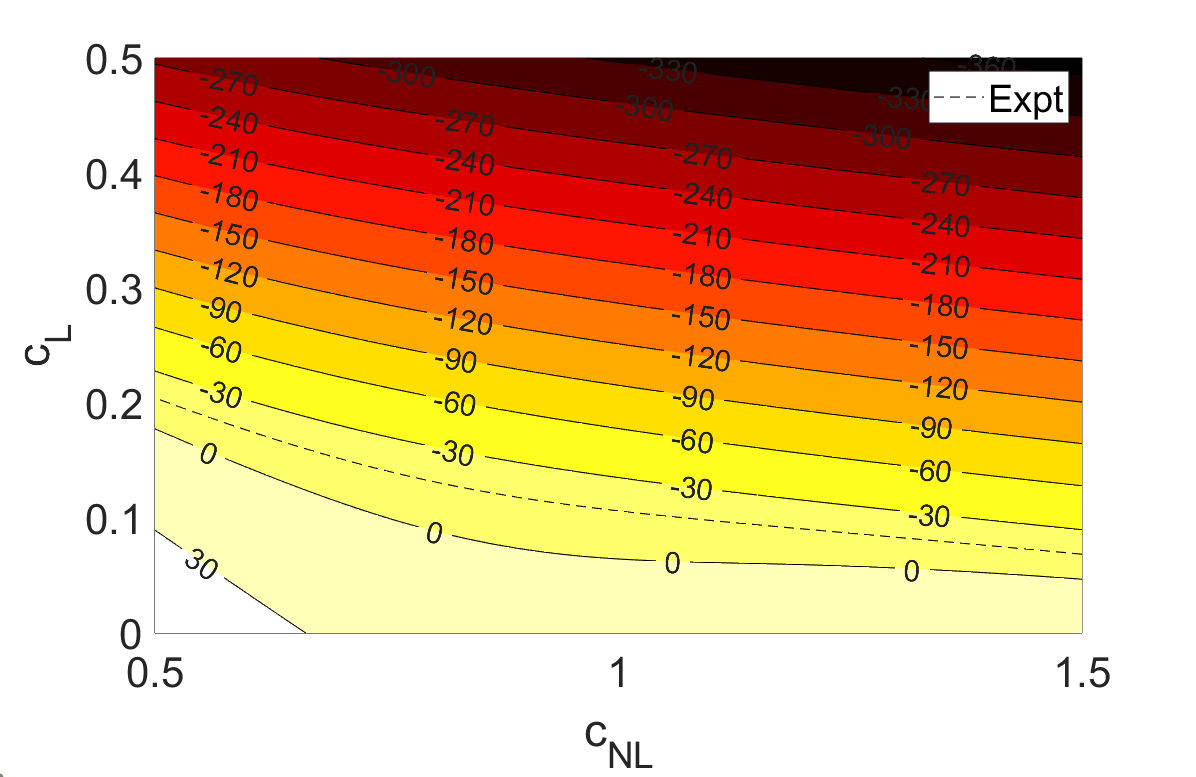} \\ 
\caption{Contour plots for the difference between the EC emulated energy for ${}^{16}$O and the four-alpha threshold energy, $E({}^{16}\text{O}) - 4E({}^{4}\text{He})$. $c_L$ is the coefficient of the local two-nucleon interaction, and $c_{NL}$ is the coefficient of the nonlocal two-nucleon interaction. 
 The dashed line shows the contour for the observed experimental value.
}
\label{fig:Emulated_contour_plot}
\end{figure}

In \textcite{Sarkar:2023qjn}, the floating block method is used to compute the binding energies of ${}^{4}$He, ${}^{8}$Be, ${}^{12}$C, and ${}^{16}$O using Monte Carlo simulations with a lattice Hamiltonian of the form $H_{\rm free} + c_LV_L + c_{NL}V_{NL}$.  $V_L$ is a two-nucleon interaction with local interactions, meaning that the interaction does not move the relative positions of the nucleons.  $V_{NL}$ is a two-nucleon interaction composed of nonlocal interactions where the relative positions of the nucleons are allowed to change.  $V_L$ and $V_{NL}$ are normalized so that $(c_L,c_{NL}) = (1,0)$ and $(c_L,c_{NL}) = (0,1)$ both give realistic results for ${}^{4}$He.  In Fig.~\ref{fig:Emulated_contour_plot}, we plot the ground state energy of ${}^{16}$O relative to the four-alpha threshold, $E({}^{16}\text{O}) - 4E({}^{4}\text{He})$ \cite{Sarkar:2023qjn}.  The EC calculation is performed with $\nb=2$ snapshots at $(c_L,c_{NL}) = (0.5,0.5)$ and $(0,1)$ in a periodic box of length $L = 15.76$~fm.  The dashed line shows the contour for the observed experimental value.  The zero contour line corresponds with the quantum phase transition where $^{16}$O falls apart into four alpha particles.  These results are consistent with the finding in \textcite{Serdar:2016} that, without sufficiently attractive local interactions, symmetric nuclear matter forms a Bose gas of alpha particles rather than a nuclear liquid.

\section{Summary and Future Directions}\label{sec:future}
In this colloquium article, we have presented the historical development, the theoretical framework, and applications of EC and projection-based emulators.  The key concept is that the eigenvector $\ket{\psi(\params)}$ is an analytic function for real values of the parameters and approximately lies on a linear subspace with a finite number of dimensions.  The smoother and more gradual the undulations, the fewer dimensions needed.  The linear subspace can be found efficiently by taking snapshots of $\ket{\psi(\params_i)}$ for the selected parameter values $\params_i$ and using the corresponding subspace spanned by the snapshots.

EC is part of a larger class of subspace projection techniques called reduced-basis methods (RBMs), and RBMs are themselves part of a yet larger category of model-driven reduced-order models.  The development of EC has emphasized applications to quantum systems, from few-body problems to many-body problems, and from bound states to scattering states and resonances.  Some of the topics addressed go beyond the traditional class of problems typically encountered in the reduced-basis literature, such as parameter extrapolation to domains that are not directly calculable, accelerating the convergence of many-body perturbation theory, and working with extremely large or infinite-dimensional vector spaces.  As noted in Sec.~\ref{sec:theory}, EC can sometimes face challenges when it is applied to situations where one tries to extrapolate across boundaries between physically distinct phases.

While the development of EC and projection-based emulators by the nuclear theory community has quite naturally focused on problems of interest for nuclear physics, e.g., uncertainty quantification and other computational statistics analyses of the nuclear Hamiltonian, the methods are quite general and can be applied to other fields where quantum wave functions are important. Remaining challenges include how to identify EC target states in the spectrum of non-Hermitian Hamiltonians and how to best handle non-affine parameter dependencies in nuclear applications. The usefulness of combining distributed emulators as mini-applications in, e.g., Bayesian inference analyses, also awaits to be capitalized on.

\begin{figure}
 \includegraphics[width=0.99
\columnwidth]{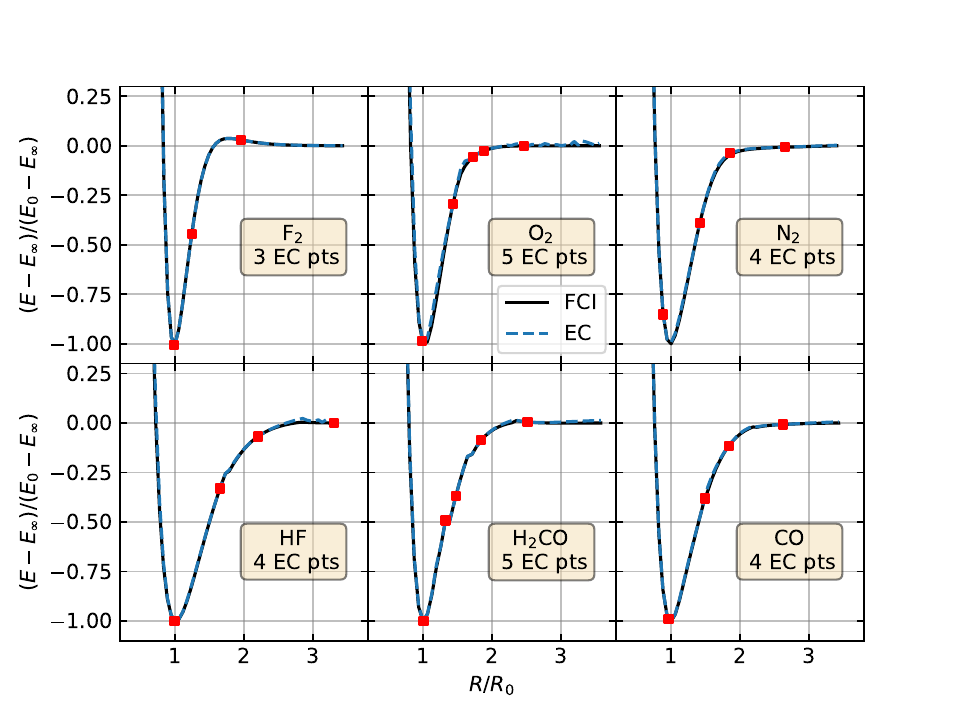}
 \caption{The potential energy versus bond stretching factor $R/R_0$ for F$_2$, O$_2$, N$_2$, HF, H$_2$CO, CO obtained using EC with several snapshots.  We show the comparison with exact FCI calculations 
 \cite{Mejuto-Zaera:2023ier}.
  \label{fig:dimers}
 }
\end{figure}

Other areas where EC and projection-based emulators should be useful are atomic and molecular physics, ultracold atomic gases, strongly-correlated electronic systems, quantum spin liquids, and quantum chemistry.  In \textcite{Mejuto-Zaera:2023ier}, EC was applied to problems in {\it ab initio} quantum chemistry.  In Fig.~\ref{fig:dimers}, we show the potential energy surface for several molecules versus bond stretching factor $R/R_0$.  The molecules are F$_2$, O$_2$, N$_2$, HF, H$_2$CO, and CO.  We show the EC results obtained with $\nb=3,4$, or $5$ snapshots and the comparison with exact FCI calculations.  Eigenvector Continuation is working well in reproducing all of the potential energy surfaces.

The potential energy surface calculations described in \textcite{Mejuto-Zaera:2023ier} can also be performed on a quantum computer, and the corresponding algorithm is called quantum EC~\cite{Francis:2022zib}.  
While much effort in the quantum computing community has focused on variational methods optimizing a single trial vector, variational calculations using subspace projection can in principle provide a better approximation to the eigenstate of interest for the same computational resources.  

In quantum EC, the same general approach is used as on a classical computer, though there are some technical differences. On a digital quantum computer with $N$ qubits, we start from the state where all qubits are in the $\ket{0}$ state, $\ket{00\cdots0}$.  Let $\ket{\psi_i}$ denote eigenstate snapshots at parameter values $\params_i$. We assume that our chosen quantum eigenstate algorithm gives us some unitary gate $U_i$ such that the action on $\ket{00\cdots0}$ gives us a good approximation to $\ket{\psi_i}$.  

The norm and Hamiltonian matrix elements can be determined using an ancilla qubit.  We use the ancilla qubit to apply the controlled operations for $U_i$ and $U^\dagger_j$.  Such a controlled operation means that we perform the transformation only if the ancilla qubit is in the $\ket{1}$ state.  After these controlled unitary operators, we then apply $\sigma_X$ or $\sigma_Y$ rotations to the ancilla qubit and measure it, with either a $\ket{0}$ or a $\ket{1}$ as the outcome.  This information is enough to determine the real and imaginary parts of $\braket{\psi_j|\psi_i}$.  In order to compute the elements $\braket{\psi_j|H(\params)|\psi_i}$, we decompose $H(\params)$ into a sum of tensor products of Pauli operators.  For each tensor product of Pauli operators, $U_{P}$, we use the ancilla qubit to apply the controlled operations for $U_i$, $U_{P}$, and $U^\dagger_j$.  We then apply $\sigma_X$ and $\sigma_Y$ rotations to the ancilla qubit and measure it \cite{Francis:2022zib}.  

EC and projection-based emulators can also be combined with data-driven reduced-order model techniques such as Gaussian processes, neural networks, and dynamic mode decomposition.  The combination of reduced-basis methods and machine learning is an active area of research and can be realized in many different ways.  As noted in Sec.~\ref{sec:MOR}, active learning methods (e.g., greedy algorithms) are often used to optimize the offline selection of snapshot parameters and projection subspaces \cite{Quarteroni:218966,Sarkar:2021fpz,chellappa2021training}.  In numerically challenging problems where convergence with respect to projection subspace dimension is slow, it is useful to treat the projection-based emulator as one component embedded within a larger framework such as a deep neural network or autoencoder \cite{brunton2019data,dal2020data,fresca2020deep,chen2021physics}. 

\begin{acknowledgements}
We are grateful to Christian Drischler, Pablo Giuliani, Gaute Hagen, and Xilin Zhang for reading the manuscript and suggesting improvements.  The work of A.E.\ is supported by the European Research Council (ERC) under the European Unions Horizon 2020 research and innovation (Grant agreement No. 758027) and the Swedish Research Council (Grant agreement No. 2020-05127).  
The work of R.F.\ is supported by the U.S.\ National Science Foundation (NSF) (PHY-2209442, OAC-2004601 -- CSSI BAND Collaboration) and the U.S.\ Department of Energy (DOE) (DE-SC0024509 
 -- STREAMLINE Collaboration, DE-SC0023175 -- SciDAC-5 NUCLEI Collaboration).  The work of S.K.\ is supported by the U.S.\ NSF (PHY-2044632) and by the U.S.\ DOE (DE-SC0024520 -- STREAMLINE Collaboration and DE-SC0024622).
This material is based upon work supported by the U.S.\ DOE, Office of Science, Office of Nuclear Physics, under the FRIB Theory Alliance, award DE-SC0013617 (S.K.).
The work of D.L.\ is supported by the U.S. NSF (PHY-2310620) and the U.S.\ DOE (DE-SC0013365, DE-SC0023658, DE-SC0024586 -- STREAMLINE Collaboration, DE-SC0023175 -- SciDAC-5 NUCLEI Collaboration), and with computational resources provided by the Oak Ridge Leadership Computing Facility through the INCITE award
``Ab-initio nuclear structure and nuclear reactions'' and the GCS Supercomputer JUWELS at the J{\"u}lich Supercomputing Centre (JSC).  
\end{acknowledgements}


\bibliography{References}
\end{document}